\documentclass{aa}  

\usepackage{graphicx}
\usepackage{txfonts}
\usepackage{mathptmx}
\usepackage{moresize}
\usepackage[percent]{overpic}
\usepackage{url}
\usepackage{comment}

\usepackage[colorlinks]{hyperref}

\hypersetup{colorlinks=blue, linkcolor=blue, urlcolor=blue, citecolor=blue}

\usepackage{graphicx}
\usepackage{physics}
\usepackage{color}
\usepackage{multirow}
\usepackage{gensymb}
\usepackage{natbib}
\bibpunct{(}{)}{;}{a}{}{,}
\usepackage[normalem]{ulem}

\begin{document}

\title{A look at the high energy aspects of the supernova remnant G309.8+00.0 with eROSITA and \textit{Fermi}-LAT}

   \author{
          M. Michailidis
          \inst{1},
           G. Pühlhofer
          \inst{1},
          A. Santangelo
          \inst{1},
          M. Sasaki
          \inst{2},
          W. Becker
          \inst{3,4}
          }

   \institute{Institut für Astronomie und Astrophysik Tübingen (IAAT), Sand 1, 72076 Tübingen, Germany   \\            \email{michailidis@astro.uni-tuebingen.de}
   \and Dr.\ Karl Remeis Observatory, Erlangen Centre for Astroparticle Physics, Friedrich-Alexander-Universit\"{a}t Erlangen-N\"{u}rnberg, Sternwartstra{\ss}e 7, 96049 Bamberg, Germany
\and Max-Planck-Institut f{\"u}r extraterrestritche Physik, Giessenbachstrasse, 85748 Garching, Germany
\and Max-Planck Institut für Radioastronomie, Auf dem Hügel 69, 53121 Bonn, Germany
             }
   \date{Received 17 April 2024 / Accepted 05 June 2024}

\abstract{Supernova remnant (SNR) detection along the Galactic plane poses a number of challenges. A diffuse X-ray emission component emanating from unidentified sources on the Galactic plane further complicates such a detection in X-rays. Due to the presence of dense dust clouds along the Galactic plane, X-ray photons are also subject to high absorption. Similarly, diffuse signals from the Galactic plane cause $\gamma$-ray contamination from the signal of individual objects. The SNR G309.8+00.0 lies exactly on the Galactic plane, with its center coinciding with galactic latitude (b)=0$\degree$. In this paper we report the first detection of the SNR G309.8+00.0 in X-rays and $\gamma$ rays, using stacked data from the first four consecutive extended ROentgen Survey Imaging Telescope Array (eROSITA) -- on board the Russian-German Spektrum Roentgen Gamma (SRG) -- all-sky surveys (eRASS:4) and $\sim15.5$~yr of Pass 8 data recorded from \textit{Fermi}-LAT, respectively. The SNR appears to have an elliptical shape of $0\degree.43\times0\degree.32$ in size in both radio synchrotron and X-ray data. The SNR's emission exhibits a shell-like morphology and good spatial correlation in both energy bands. The X-ray emission was solely detected in the 1-2 keV energy band (subject to strong absorption at soft X-rays) and the spectral analysis results of eRASS:4 data present a purely thermal SNR with a high absorption column density $3.1_{-0.5}^{+0.7}\cdot10^{22}~\mathrm{cm^{-2}}$ and a temperature of $0.34\pm0.1$~keV. Although the thermal plasma appears to be in equilibrium, the limited statistics do not allow us to exclude nonequilibrium models. 
The X-ray spectral analysis of the remnant resulted in the detection of relatively (given the limited statistics) prominent Mg triplet lines at 1.33-1.47~keV and silicon (Si XIII) at 1.74-1.9~keV energies.  
In combination with optical extinction data, the absorption column density values derived from the remnant's spectral analysis support a remnant's distance greater than 6~kpc, rather than a 3.12~kpc distance as reported in the literature, and yield an age of $1-3.5\cdot10^5$~yr. Employing $\sim15.5$~yr of \textit{Fermi}-LAT $\gamma$-ray data at and around the remnant's vicinity, we confirm the detection of the to-date unidentified 4FGL J1349.5-6206c source that can either be modeled as a single source or a conglomerate of multiple distinct source components. In the latter case, the detailed inspection of the \textit{Fermi}-LAT $\gamma$-ray data in the direction of the remnant allowed us to decompose the 4FGL J1349.5-6206c source into four point-like components, among which one is spatially coincident with the SNR G309.8+00.0 shell.
We detected the component that spatially coincides with the SNR with a significance of $5.8\sigma$ above $1$~GeV with \textit{Fermi}-LAT and thus argue that the SNR G309.8+00.0 likely represents at least a significant portion (if not all) of the emission from the 4FGL J1349.5-6206c $\gamma$-ray source, detected with $9.8\sigma$ significance $>1$~GeV with \textit{Fermi}-LAT.

}


\titlerunning{A look at the high energy aspects of the SNR G309.8+00.0 with eROSITA and \textit{Fermi}-LAT}
\authorrunning{M. Michailidis}
\maketitle

\section{Introduction}
\label{sec:intro}
For the last two decades, Galactic supernova remnants (SNRs) have been confirmed to accelerate particles (electrons and nuclei) up to the highest energies. The first detection of X-ray synchrotron emission from SNRs was reported by \citet{1995Natur.378..255K}, and since then, several tens of Galactic SNRs have been detected to be purely nonthermal in X-rays or to exhibit at least a nonthermal component in their X-ray spectrum. In the last decade, a few tens of Galactic SNRs have been detected at even higher energies, that is, gigaelectronvolt and teraelectronvolt energies \citep{2016ApJS..224....8A,2018A&A...612A...1H,2018A&A...612A...3H}. Accelerated teraelectronvolt electrons may emit X-rays through synchrotron \citep{1995Natur.378..255K}, but may also emit $\gamma$ rays through inverse Compton (IC). In addition, in a number of $\gamma$-ray SNRs, the characteristic pion-decay signature has been detected, providing evidence for proton (nuclei) acceleration from SNRs \citep{2013Sci...339..807A}.
Lately, there has been ample evidence, for example \citet{2024arXiv240117312M,2024A&A...685A..23M} and \citet{2024arXiv240117261K}, 
that an increasing number of relatively old SNRs observed in $\gamma$ rays, of hadronic origin, are characterized by a purely thermal emission component in X-rays and the absence of a nonthermal X-ray component. Since the same population of relativistic teraelectronvolt electrons is responsible for both X-ray synchrotron emission and $\gamma$-ray emission of leptonic origin, it is not surprising that the latter assertion holds, for example, readers can refer to HESS J1614-518 (P\"uhlhofer et al., in prep.), a newly identified SNR with a nonthermal X-ray spectrum accompanied by $\gamma$-ray emission of leptonic origin. However, no such Universal correlation has been confirmed across all $\gamma$-ray SNRs. The total number of detected Galactic SNRs is $\sim300$ -- Green catalog \citep{2019JApA...40...36G}, SNRcat\footnote{\url{http://snrcat.physics.umanitoba.ca/}} \citep{2012AdSpR..49.1313F}. Although only a small fraction, on the order of $\sim10\%$, of Galactic SNRs are observable in $\gamma$ rays to date, the study of those objects in the highest energies and the detection and identification of new SNRs emitting in $\gamma$ rays is of great importance to gain further insight into the particle acceleration in our Galaxy and the fraction of the energy budget of the cosmic-ray (CR) spectrum that is attributed to those objects.

X-ray and $\gamma$-ray emission from Galactic SNRs is difficult to detect when the latter objects lie along the Galactic plane. The reason for this is that X-ray photons are strongly absorbed due to the prevalence of dust clouds on the Galactic plane. At the same time, the X-ray and $\gamma$-ray signals from the Galactic plane are strongly contaminated by diffuse emission, in both energy bands, potentially originating from dozens of nearby unidentified objects. Additionally, X-ray and $\gamma$-ray emission is only observed at specific stages of the SNR evolution posing additional difficulty to detecting SNRs in the latter energies from the entire population of observed Galactic SNRs.

On the contrary, the fact that radio synchrotron emission, which is not subjected to strong absorption, stems from a population of gigaelectronvolt electrons that lose energy slower compared to higher energy particles (i.e., teraelectronvolt electrons responsible for X-ray synchrotron emission) makes it an ideal energy range to search for Galactic SNRs. In fact, the large majority of Galactic SNRs are observed in radio given that the lifetime of the responsible particles for the emission is greater than the age of the SNRs. Consequently, the main criterion for the classification of an object as a SNR is the detection of a shell-type morphology and a nonthermal spectral index in radio synchrotron data.
However, an increasing number of Galactic SNRs has been detected at higher energies -- X-rays and $\gamma$ rays -- in recent years with the latest X-ray all-sky surveys: now the extended ROentgen Survey Imaging Telescope Array (eROSITA) and previously the ROentgen SATellite (ROSAT) (detection of several tens of Galactic SNRs emitting in X-rays), the \textit{Fermi}-LAT $\gamma$-ray all-sky survey \citep{2016ApJS..224....8A}, and also with H.E.S.S. at teraelectronvolt energies \citep{2018A&A...612A...3H,2018A&A...612A...8H}. 
By investigating those objects in X-rays and $\gamma$ rays, we are able to gain valuable information about their nature, including the physics of shocks, magnetic field strength, heating, and acceleration mechanisms, as well as individual properties such as their distance and age estimates.

In this work, we report the first detection of the SNR G309.8+00.0 in X-rays and the identification of at least a significant fraction of the emission from the to-date named 4FGL J1349.5-6206c \textit{Fermi}-LAT source as the remnant's gigaelectronvolt counterpart by utilizing stacked data from the first four consecutive eROSITA all-sky surveys (eRASS:4) and $\sim15.5$~yr of Pass 8 data recorded from \textit{Fermi}-LAT, respectively. With its center lying on the Galactic plane, G309.8+00.0 belongs to the class of SNRs that are subjected to strong absorption features at softer X-rays ($<1$~keV) mainly due to its location. The remnant was detected for the first time in the radio wavelengths as an elliptical shell and identified as a SNR due to its nonthermal spectral index with the high resolution surface brightness contour maps at 480~MHz and 5000~MHz obtained by the Molonglo cross telescope and the Parkes 64-m radio telescope, respectively \citep{1975AuJPA..37....1C}. A strong radio point source was also detected nearly at the center of the shell, just 4' to the north of the remnant's geometrical center. However, due to its steep spectrum (according to the Molonglo Observatory Synthesis Telescope (MOST) data, an -1.0 spectral index was derived  for the central point source \citep{1996A&AS..118..329W}, a value which is consistent with background extragalactic sources) and lack of pulsations, it is considered to be quite likely of extragalactic origin. The latter point-like object appears to have an optical counterpart, currently named 2MASS J13503303-6159245 or Gaia DR3 5865571591306465280, however, due to its low galactic latitude (b) the spatial coincidence might occur by chance. Spatial and spectral consistent results, compared to the above studies, were obtained with higher resolution maps obtained from the Fleurs synthesis telescope at 1415~MHz \citep{1980MNRAS.190..881C}. However, the first detailed grayscale map, rather than contours, was reported by \citet{1996A&AS..118..329W} by exploiting MOST radio data at 843~MHz. In the latter maps, the remnant appears as a well-defined shell  
of elliptical shape and $25'\times19'$ size. Finally, no masser (OH) detection from the satellite line of the hydroxyl radical (OH) at 1720.5 MHz with Parkes telescope was found, as reported in \citet{1997AJ....114.2058G}. 

A recent study by \citet{2020A&A...639A..72W} on the SNRs distances in the inner disk, utilizing red clump stars (RCS) as tracers, resulted in a $3.12\pm0.22$~kpc primary distance estimate with the highest reliability. However, the authors note that this SNR, as several others in their study, exhibits two distance gradients. The secondary distance measurement is estimated to be $5.61\pm0.42$~kpc. Finally, \citet{2013ApJS..204....4P} reported a 4~kpc remnant's distance based on the radio surface-brightness-to-diameter distance estimate method.

Except for an unidentified \textit{Fermi}-LAT gigaelectronvolt source, 4FGL J1349.5-6206c, that could potentially be associated with the SNR mainly due to its location; the SNR G309.8+00.0 has not been detected in any other wavelengths. We note that the latter $\gamma$-ray source has not been studied individually and has been modeled as a point-like object in the latest \textit{Fermi}-LAT catalogs (4FGL-DR4, 4FGL-DR3 \citep{2023arXiv230712546B,2022ApJS..260...53A}). It is noteworthy that even though the latter source was initially included in the 1FGL catalog \citep{2010ApJS..188..405A}, it was then removed from the 2FGL \citep{2012ApJS..199...31N} and 3FGL catalogs \citep{2015ApJS..218...23A}, and it was not included in the first \textit{Fermi}-LAT catalog of SNRs \citep{2016ApJS..224....8A}. A weak signal of 1.3$\sigma$ significance was reported from the remnant's location as part of a population study of Galactic SNRs at very high energies (teraelectronvolts) with H.E.S.S. \citep{2018A&A...612A...3H}.

This paper is outlined as follows. Section~\ref{sec:intro1} is dedicated to the X-ray imaging and spectral analysis of the SNR G309.8+00.0 using eRASS:4 data. In section~\ref{sec:intro2} we report on the distance and age estimates of the remnant based on the absorption column density values derived from the X-ray spectral fitting. In addition, we examine potential pulsar associations with the remnant. Section~\ref{sec:intro3} describes the detailed inspection of the spatial morphology of the $\gamma$-ray emission detected at and around the remnant with \textit{Fermi}-LAT (4FGL J1349.5-6206c). The gigaelectronvolt spectral energy distribution (SED) for the 4FGL J1349.5-6206c and its component which is spatially coincident with the SNR G309.8+00.0 is also presented by putting the measured fluxes into context. Section~\ref{sec:intro4} gives concluding remarks.

\begin{figure*}
    \sidecaption
    \includegraphics[width=12cm,clip=true, trim= 0.1cm 0.8cm 2.5cm 2.7cm]{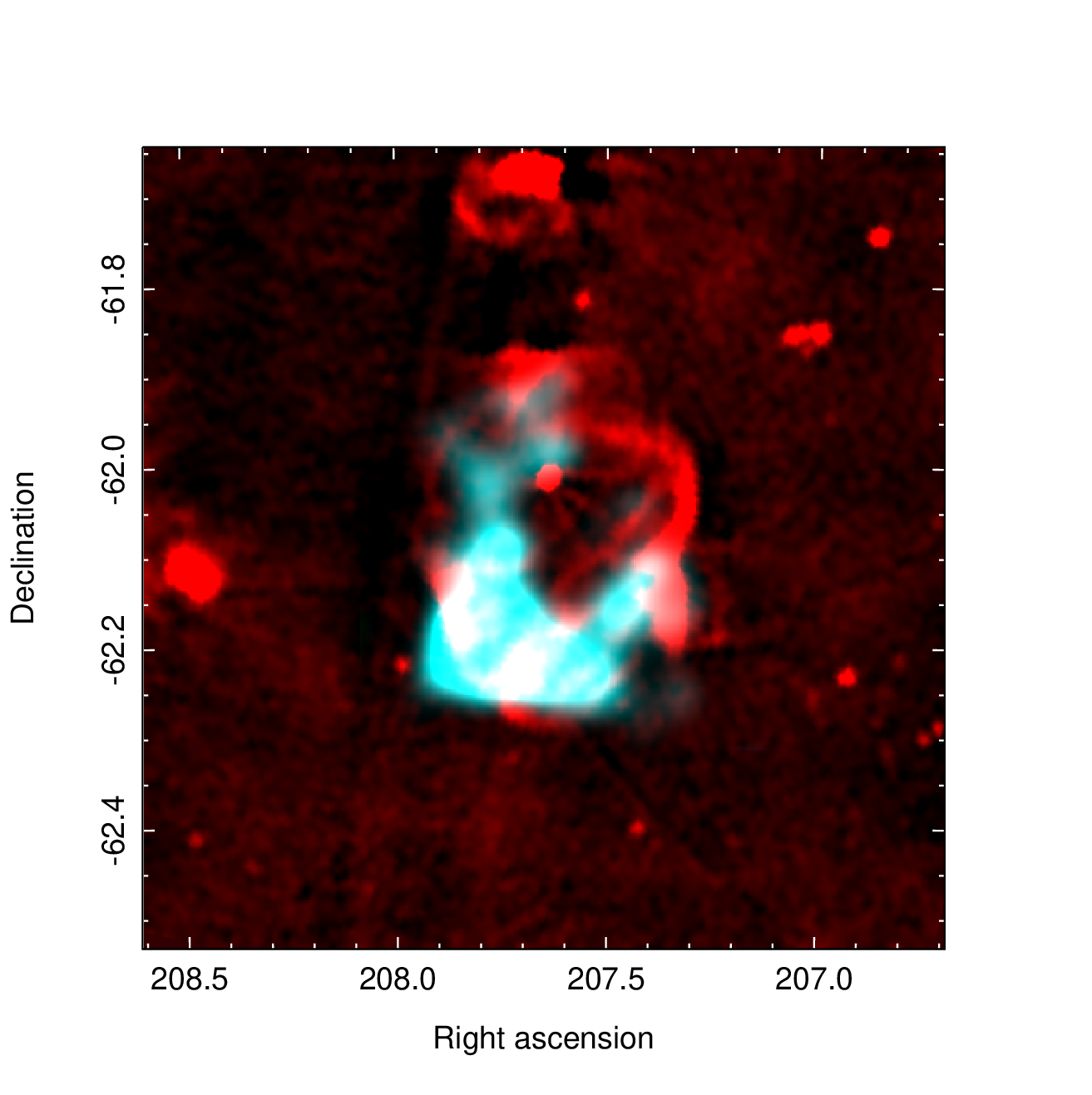}
    \caption{
    Combined SUMSS at 843 MHz (red) and 1-2 keV eRASS:4 (cyan) image at and around the SNR G309.8+00.0 location. The X-ray emission component is restricted to the 1.0-2.0\,keV energy band in which the X-ray emission from the remnant is concentrated. A linear color distribution is used for both image components. The image components are plotted in units of Janskys per beam and counts per pixel with a pixel size of $11''$ and $4''$ for the radio and X-ray components, respectively. The radio component is convolved with a $1\sigma$ Gaussian kernel and the X-ray component has been adaptively smoothed with a Gaussian kernel of $5\sigma$ to enhance the visibility of the diffuse emission. Point sources above $3\sigma$ significance are filtered out from the X-ray image using a $60''$ radius kernel.
    }
    \label{mainimage}
\end{figure*}
\section{X-ray data analysis}
\label{sec:intro1}

In this work, we report on the analysis results of X-ray data taken during the first four eROSITA all-sky surveys (eRASS:4), in the c020 processing version. eROSITA is the primary instrument aboard the Russian-German Spektrum Roentgen Gamma (SRG) observatory \citep{2021A&A...656A.132S}. It consists of seven parallel aligned telescopes of $1\degree$ field of view (FoV) each (TM1-7) operating in the 0.2-10.0~keV energy range \citep{2012arXiv1209.3114M,2021A&A...647A...1P}. It achieves a $\sim30''$ average spatial resolution in survey mode \citep{Merloni2023}. In contrast to the rest of the telescopes, modules TM5 and TM7 are not equipped with an aluminum on-chip filter, and thus to avoid signal contamination that could distort further estimates carried out below, mainly due to potential light leak suffering \citep{2021A&A...647A...1P}, data collected from TM5 and TM7 were used for imaging analysis purposes but were excluded in the spectral analysis process. $\texttt{evtool}$ and $\texttt{srctool}$ tasks of the eROSITA Standard Analysis Software (eSASS) version $\texttt{eSASSusers\_201009}$ \citep{2022A&A...661A...1B} were employed for data reduction and data processing purposes. From the 4700 partially overlapping sky tiles (of $3\degree.6\times3\degree.6$ size each) of eSASS pipeline, the SNR G309.8+00.0 lies entirely on the 205153 tile. Each tile is named after its center coordinates. The first three digits correspond to the RA whereas the last three to the Dec of the tile's center. Only data from the sky tile mentioned above were used in this work.
\begin{figure*}[h!]
    \sidecaption
    \includegraphics[width=12cm,clip=true, trim= 0.1cm 0.2cm 1.35cm 1.3cm]{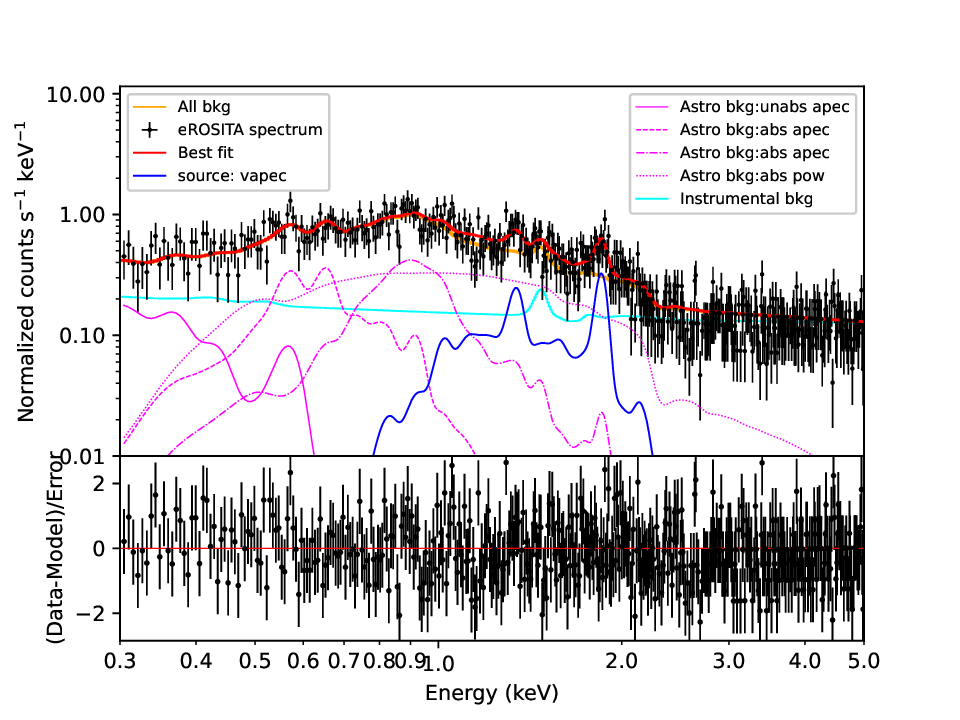}
    \caption{Energy spectrum in the 0.3-5.0~keV energy band obtained from the selected on-source region as defined in the text (section~\ref{sec:intro2}), using eRASS:4 data obtained from TM1,2,3,4, and 6. Following the fitting procedure, a rebinning process that combines two spectral bins into one was used solely for visual purposes. The G309.8+00.0 X-ray energy spectrum is best described by a thermal plasma component in equilibrium (tbabs$\times$vapec) indicated with the blue solid line.
    All components contributing to the total emission are displayed in detail. The orange line stands for the sum of all background components, that is, instrumental (cyan color), and astrophysical (magenta color), detected at and around G309.8+00.0, revealing the overall contribution of the source to the total spectrum.}
    \label{eRASSspec}
\end{figure*}

\subsection{First X-ray detection and X-ray imaging analysis}\label{Xima}

The combined Sydney University Molonglo Sky Survey (SUMSS) at 843 MHz (red) and 1-2 keV eRASS:4 (cyan) image of the SNR G309.8+00.0 is shown in Fig.~\ref{mainimage}. The X-ray image component (intensity sky map) was optimized aiming at emphasizing the diffuse X-ray emission originating from the SNR and avoiding likely contamination of the signal from unrelated nearby sources. All X-ray point-like sources with at least a 3$\sigma$ detection significance were masked out using a kernel radius of $60''$. The X-ray image, of $4''$ pixel size, has been adaptively smoothed according to the smoothing algorithm of \citet{10.1111/j.1365-2966.2006.10135.x} with a Gaussian kernel of $5\sigma$ to enhance the diffuse X-ray emission visibility. The majority of the X-ray emission emanating from the SNR is confined in the 1-2~keV energy band. The latter energy range was optimized after detailed inspection of the remnant's X-ray spectrum. The emission in both radio synchrotron and X-rays appears to be in good morphological correlation and confined within a well-defined shell of elliptical shape. However, significant X-ray emission is only detected in the lower half of the shell. As shown by the X-ray imaging analysis, the northern half and in particular the northern edge of the remnant's shell are only marginally detected above the background level, $<3\sigma$ detection significance in X-rays. Additionally, G309.8+00.0 can be classified as a limb-brightened SNR, similar to, for example, the Cygnus loop SNR. We observe a somewhat incomplete ring-like structure in X-rays since there is more hot gas in our line of sight at the edges compared to when looking through the central parts of the remnant. However, the limited statistics of the data forbid the construction of a high-quality image that demonstrates such a comparison between the brightness of the remnant at the central regions compared to the edges since the central regions brightness is only marginally above the background level. A size of $0\degree.43\times0\degree.32$ is obtained in X-rays, by fitting an ellipse to the outermost regions of emission, in excellent agreement with the remnant's radio size reported in the literature.

\subsection{X-ray spectral analysis}\label{Xspect}

The $\texttt{srctool}$ task of the eSASS pipeline was used to extract the corresponding spectral files for spectral fitting purposes. The collected data from TM5 and TM7 were excluded to avoid contamination issues, as mentioned in section~\ref{sec:intro1}. The X-ray spectral analysis was carried out using the to-date version of the Xspec code \citep{Borkowski_2006,10.1093/pasj/63.sp3.S837}. Given the faint appearance of the SNR in X-rays ($300^{+210}_{-120}$ source counts in 1-2~keV energy range, depending on the selected background control region), Cash statistics \citep{1979ApJ...228..939C} was preferred over Chi-Square \citep{Pearson1900} for the evaluation of the goodness of the fit and no rebinning of the X-ray photons was applied prior to the fitting process. No significant X-ray point sources were detected within the remnant's extension. However, an identical strategy to section~\ref{Xima} was adopted to avoid potential signal contribution to the remnant's X-ray spectrum by unrelated point sources. Points sources above 3$\sigma$ significance were masked carefully with a kernel radius of $60''$ since a larger masking radius would remove a significant fraction from the faint diffuse X-ray emission of the remnant. 

Table~\ref{fit} summarizes the spectral parameters and 1$\sigma$ error results of the best-fit model to the X-ray spectrum of the SNR G309.8+00.0 (shown in Fig.~\ref{eRASSspec}). The latter results were obtained by performing a simultaneous fitting of the source and background emission from the selected on-source region (a circular region centered at (l)=$309.785\degree$, (b)=$-0.025\degree$ with a radius of $0.27\degree$). The parameters of the background model, used to describe the background emission from the on-source region, were fixed to the best-fit values obtained from the spectral analysis of a surrounding background control region located to the southeast of the remnant that is free of source emission (a circular region centered at (l)=$310.11\degree$, (b)=$-0.4\degree$ with a radius of $0.27\degree$). No rescaling was required since the on-source and background control regions were selected to be of the same area. The latter background control region was selected to represent best the background emission of the on-source region after carefully inspecting the surrounding regions for potential contamination from foreground and/or background sources. The background emission was found to be best described by the following model in Xspec notation: \texttt{apec+tbabs(apec+apec+pow) + gaussian + expfac(bkn2pow + powerlaw + powerlaw) + powerlaw + gaussian + gaussian + gaussian+ gaussian + gaussian + gaussian + gaussian + gaussian + gaussian}. The latter model is a convolution of the i) astrophysical background (\texttt{apec+tbabs(apec+apec+pow)}) which represents the Local Hot Bubble (LHB) low temperature plasma, the Galactic Halo (GH) plasma, and the Cosmic X-ray Background (CXB) -- the product of the combined emission from unresolved Active Galactic Nuclei (AGN), and ii) of the instrumental background which can be expressed as a combination of power law and Gaussian model components in the selected energy range for spectral fitting: \texttt{gaussian + expfac(bkn2pow + powerlaw + powerlaw) + powerlaw + gaussian + gaussian + gaussian+ gaussian + gaussian + gaussian}.

\begin{table}[!]
    \caption{Best-fit parameters of the SNR G309.8+00.0 X-ray spectrum.} 
    \label{fit}
    \centering
    \begin{tabular}{c|c}
    \hline 
        \textbf{Model} & \textbf{tbabs$\times$vapec}    \\
        \hline
        kT(keV)&  $0.34_{-0.1}^{+0.1}$ \\\hline
       $N_{H}(10^{22}~\mathrm{cm^{-2}})$ &$3.1_{-0.5}^{+0.7}$ \\\hline
        $\mathrm{Si/Si_{\odot}}$ &$3.6_{-1.2}^{+2.3}$  \\
        \hline
        norm &$0.08^{+0.18}_{-0.04}$\\\hline
        CSTAT/d.o.f &0.96\\\hline
        \end{tabular}
        \tablefoot{The best-fit parameters are shown along with $1\sigma$ errors. Where not defined otherwise, elemental abundances are set to nearly solar values according to \citet{2000ApJ...542..914W}.}
        
\end{table}
\begin{figure}
    \centering
    \includegraphics[width=0.5\textwidth]{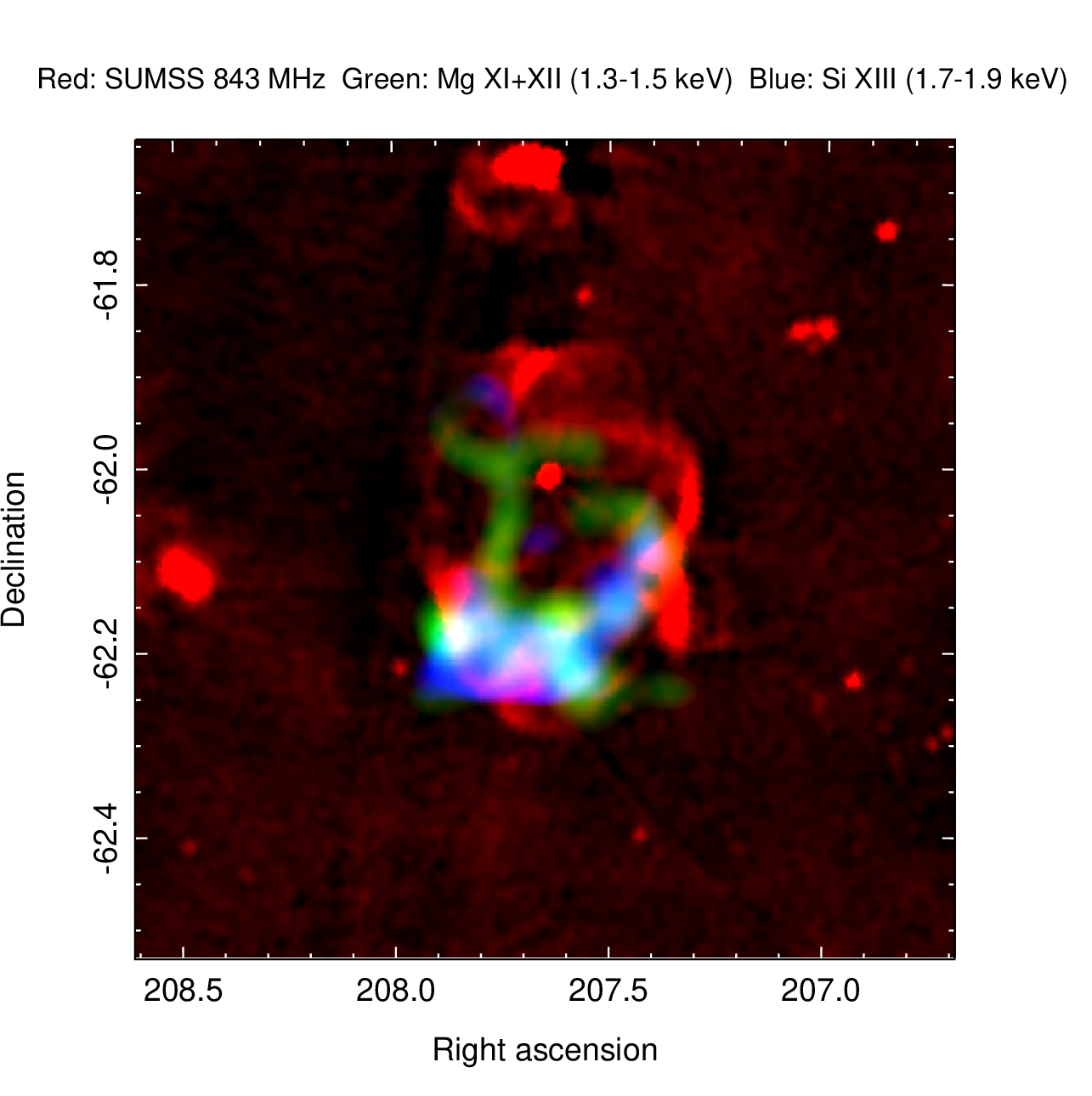}
    \caption{
    Combined SUMSS at 843 MHz (red), 1.3-1.5 keV eRASS:4 (green), and 1.7-1.9 keV eRASS:4 (blue) image at and around the SNR G309.8+00.0 location. A spectrally motivated selection of X-ray narrow energy ranges was made to demonstrate the spatial distribution of Mg XI+XII and Si XIII emission lines across the remnant's area. A linear color distribution is used for all three image components. The radio and X-ray components of this sky map were constructed using an identical approach in terms of pixel size, smoothing, and point source removal as described in the caption of Fig.~\ref{mainimage}.
    }
    \label{secondimage}
\end{figure}
The X-ray emission originating from the SNR appears to be purely thermal. A simple absorbed thermal plasma in equilibrium appears to describe its spectrum well ($\mathrm{tbabs\times vapec}$ as described in Table~\ref{fit}). Nonequilibrium thermal components were tested and provide fits of equal goodness with similar best-fit spectral parameters (e.g., for a $\mathrm{tbabs\times vnei}$ model (almost identical parameters were obtained for a $\mathrm{tbabs\times vpshock}$ model), kT=  $0.34_{-0.13}^{+0.13}$~keV, $N_{\mathrm{H}}=3.4_{-0.8}^{+1.0}~10^{22}~\mathrm{cm^{-2}}$,  $\mathrm{Si/Si_{\odot}}=2.6_{-0.8}^{+4.5}$, and $\tau=1.42^{+unconstrained}_{-1.07}\cdot10^{11}~\mathrm{cm^{-3}~s}$ -- unconstrained, with 1$\sigma$ errors). The derived ionization timescales ($\tau>10^{11}~\mathrm{cm^{-3}~s}$ -- full ionization equilibrium is typically reached at $\tau$ values $\geq10^{12}~\mathrm{cm^3 s}$ 
 \citep{1984Ap&SS..98..367M}) support the hypothesis for a plasma that broadly speaking is found in equilibrium. We note, however, that the limited statistics of the data do not allow us to reject nonequilibrium models (i.e., $\mathrm{tbabs\times vpshock}$ and $\mathrm{tbabs\times vnei}$ \citep{2001ApJ...548..820B,2000ApJ...542..914W}). In addition, the limited photon statistics pose a significant challenge in assessing in detail the remnant's X-ray spectrum. They do not permit us to examine the X-ray emission from individual subregions such as the most prominent arc at the southeast of the SNR which also spatially coincides with the brightest arc of its radio synchrotron emission. Thus, given the size of the SNR we plan to request a deep follow-up pointed observation to provide a more robust spectral classification of the remnant. It should be noted that an \textit{XMM-Newton} observation, ObsId: 0742110101, which was conducted to study X-ray emission from the nearby radio quiet SNR G309.4-0.1, is marginally covering the remnant, although the extremely small overlap region, which does not exhibit enhanced X-ray emission, prohibits further investigation.  Despite the faint appearance of the remnant, a relatively (given the limited statistics of the X-ray data) prominent Mg triplet line at 1.33-1.47~keV and Si emission (XIII He$\alpha$ lines) at 1.7-1.9~keV were identified. Driven by the physics conclusions obtained from the X-ray spectral analysis of the remnant, we constructed an RGB image with spectrally motivated energy to color correspondence (Red: SUMSS 843~MHz, Green: eRASS:4 1.3-1.5~keV, Blue: eRASS:4 1.7-1.9~keV), as shown in Fig.~\ref{secondimage}. The latter figure describes how the two most prominent elemental abundances distribute across the remnant's area. The Si emission appears to be concentrated mainly in the southeastern part of the shell, whereas the Mg emission appears to be homogeneously distributed on a ring-like morphology that follows the edges of the remnant's elliptical shape.

By putting the detected X-ray photons into context, we derived a surface brightness of $7.32\cdot10^{-4}~\mathrm{photons~arcsec^{-2}}$, estimated from an area of $2.93\cdot10^6~\mathrm{arcsec^2}$ -- on-source region -- and a total flux of $F_{\mathrm{total}}=1.42_{-0.51}^{+0.87}\cdot 10^{-11}~\mathrm{erg~cm^{-2}~s^{-1}}$ (1$\sigma$ errors) in the 1-2~keV energy range. Thus, we argue that the SNR G309.8+00.0 can be classified among the faintest Galactic SNRs ever detected in X-rays.

\section{Distance and age estimates}
\label{sec:intro2}
Three distance estimates have been proposed to date for the SNR G309.8+00.0, which indicate that the SNR is probably located at a distance $<4$~kpc, as discussed in detail in section~\ref{sec:intro}. 
\begin{figure}[h!]
    \centering
    \includegraphics[width=0.5\textwidth]{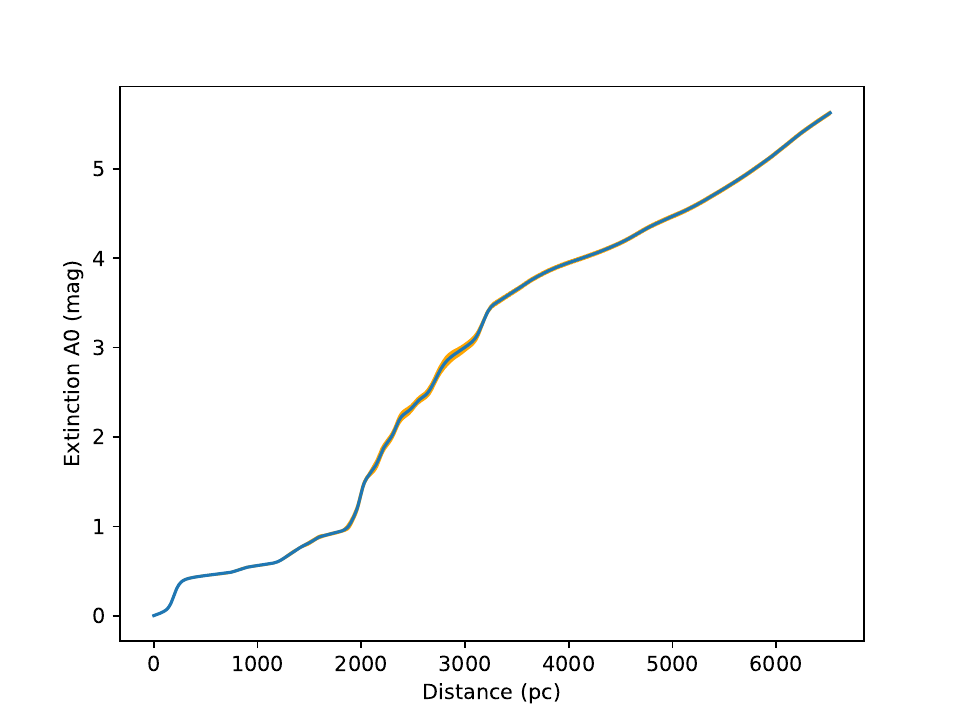}
    \caption{One-dimensional cumulative extinction as a function of the distance in the direction of the SNR G309.8+00.0. The graph was created by using the EXPLORE G-Tomo app \protect\url{https://explore-platform.eu/} which provides updated extinction data sets extended up to $\sim6$~kpc by using GAIA eDR3 and 2MASS data \citep{2022A&A...661A.147L}. The orange range indicates the area of uncertainty where significant uncertainty exists.}
    \label{ext}
\end{figure}
\begin{table*}[!]
    \centering
    \caption{Pulsars within $2\degree$ of the remnant's center.}
    \label{psr}
    \begin{tabular}{c|c|c|c|c|c|c}
    \hline\hline 
        \textbf{Pulsar} & Ang. sep.  & DM&\textbf{$D_1$} &$D_2$& Age & $v_{\mathrm{transv}}$  \\
         & ($\degree$) & \textbf{$\mathrm{pc\cdot cm^{-3}}$} & kpc&kpc& Myr &$\mathrm{km\cdot s^{-1}}$\\
        \hline
        J1337-6306& $1.84$ & $777.7$ & $12.9 (14.1)$ &- &$9.3$ &43.6 \\
        J1338-6204 & $1.46$ & $640.3$ & $12.4 (9.9)$ &- &$1.4$ &220.7 \\
        J1341-6220 (4FGL J1341.7-6216)& $1.07$&$719.7$  & $12.6 (11.2)$  &- &$0.01$&$2.3\cdot10^4$ \\
        J1344-6059& $1.3$ & $435$ & $7.4 (7.6)$ & - &$200$&0.8  \\
        J1345-6115 & $1.01$ & $278$ & $5.6 (5.0)$ &-& $6.1$ &15.8 \\
        J1348-6307 & $1.06$ & $597$ & $10.6 (10.3)$&- &$3.9$ &49.2\\
        J1349-6130& $0.59$ & $283.9$ & $5.5 (5.0)$&- &$0.8$ &69.2 \\
        J1349-63 & $1.86$ & $478$ & $9.3 (9.5)$&- &$-$ &$2.95\cdot10^4$/$2.95\cdot10^3$(*) \\
        J1350-6225 (4FGL J1350.6-6224) & $0.35$ & $-$ & $-$&-& $0.25$ &74.5/239($\dagger$) \\
        J1354-6249 & $0.87$ & $254$ & $5.2 (4.6)$&-&$3.2$ &24.1 \\ 
        J1355-6206 & $0.55$ & $547$ & $7.5 (8.3)$&-&$1410$ &0.05 \\
        J1357-62& $0.88$ & $416.7$ & $6.5 (6.7)$&-&$-$ &$9.8\cdot10^3$/980(*) \\
        J1358-6025 (4FGL J1358.3-6026) & $1.88$ & $-$ & $-$&-&$0.32$ &312.8/1003($\dagger$) \\
        J1359-6038 & $1.83$ & $293.7$ & $5.5 (5.2)$&-&$0.32$ &536.7 \\
        J1400-6325 & $1.78$ & $563$ & $9.2 (11.3)$& $7.0$&$0.01$ &$2.8\cdot10^4$ \\
        J1403-6310 & $1.82$ & $305$ & $5.7 (5.5)$&-&$70.3$ &2.52 \\\hline
        \end{tabular}
        \tablefoot{The first and second columns provide the pulsar's name and angular separation from the remnant's center. The third column gives the dispersion measure (DM). The fourth and fifth columns present the pulsar's distance from Earth based on DM measurements and potential associations, respectively. The values within parentheses correspond to older distance estimates based on the NE2001 electron density model \citep{cordes2003ne2001i}. Since 2017, the electron density model for DM-based distance calculations has been substantially updated to the current version YMW16 \citep{2017ApJ...835...29Y}. The sixth column gives the pulsar's spin-down age. The seventh column displays the transverse velocity required for each pulsar to move from the remnant's center to its present location. No proper motion estimates are provided for all the above pulsars. (*)$v_{\mathrm{transv}}$ estimate based on the SNR age $10^4/10^5$~yr. ($\dagger$) $v_{\mathrm{transv}}$ estimate based on the two distinct SNR distance approaches considered in this work $3.12/10$~kpc.}
\end{table*}
In this work, we used eRASS:4 data aiming to provide the first distance estimate based on the properties of the remnant itself.
Making use of the absorption column density values obtained from the X-ray spectral fit and adopting the most recent statistical relation between X-ray absorption and mean color excess \citep{2016ApJ...826...66F} based on \textit{Chandra} observations of SNRs

\begin{equation}
  \begin{array}{lcl}
       N_\mathrm{H}/E_{\mathrm{B-V}}&=&8.9\times10^{21}~\mathrm{cm^{-2}\cdot mag^{-1}}\\N_\mathrm{H}\mathrm{[cm^{-2}}/A_{\mathrm{\nu}}]&=&2.87(\pm0.12)\times10^{21}
 \label{math1}
 \end{array}
,\end{equation}

an optical extinction of $A_{\nu}\equiv A_0\equiv A(550~\mathrm{nm})=10.7_{-2.1}^{+2.7}$ is derived (errors were estimated by taking into account
errors in the absorption column density and errors defined from the empirical relation used). Employing the latest optical extinction data sets by \citet{2022A&A...661A.147L}, a distance well above 6~kpc (on the order of 10~kpc) is derived, as shown in Fig.~\ref{ext}. These distance values are inconsistent with earlier estimations of the remnant's distance reported in the literature. However, all distance method estimates for this remnant are subject to large scattering and uncertainty. Therefore, we discuss below the linear size and age estimates of the remnant adopting two distinct distances suggested as best from this work and the literature.

By adopting a distance to the remnant of 3.12~kpc and 10~kpc, and taking into account the remnant's angular size of $0\degree.43\times0\degree.32$, one derives a $23.3\times17.3$~pc and $75\times56$~pc size, respectively. Using the absorption column density obtained from the X-ray spectral fit, one derives a local density of $3.2^{+0.9}_{-0.7}~\mathrm{cm^{-3}}$ and $1.0^{+0.2}_{-0.2}~\mathrm{cm^{-3}}$, respectively (errors were estimated by taking into account
errors in the absorption column density). Employing the SNR evolutionary model calculator developed by \citet{Leahy_2017}, one obtains a $0.9-2.0\cdot10^4$~yr and a $1.0-3.5\cdot10^5$~yr remnant's age (errors were estimated by taking into account
errors in the computed local density) for the two distinct cases. We note that such age estimates are only rough approximations especially due to the limited statistics of the collected X-ray photons. It is, however, not possible to obtain a more accurate distance measurement since the SNR does not have a firm association with nearby pulsars (in fact, if future studies probe a type Ia progenitor origin; the remnant is not considered to be associated with a pulsar), and as discussed in section~\ref{sec:intro3}, there are no apparent molecular clouds associated with the remnant.

To this end, we also checked for potential pulsars in the near vicinity of the SNR that could be possibly associated with the remnant if the latter is not of type Ia progenitor origin. There are 16 nearby pulsars within $2\degree$ from the remnant's center, as shown in Table~\ref{psr}. Among those, nine are highly unlikely to be associated with the SNR due to their old age, which exceeds $10^6$~yr. Among the rest, two pulsars do not have computed ages and thus cannot be examined in detail (J1349-63, J1357-62). However, assuming the remnant's age on the order of $10^4-10^5$~yr (within uncertainties from the two different approaches employed above), the extremely high pulsars' transverse velocities required to reach their present location do not permit a remnant's association with those two pulsars. Among the five remaining pulsars with compatible age with the SNR, only two have acceptable transverse velocities. Since for those two pulsars there is no distance estimate, we exploited both possible distance estimates of the remnant adopted in this work. The J1350-6225 (4FGL J1350.6-6224) pulsar, which is located just $0.35\degree$ away from the remnant's center (still outside the remnant's extension), has an age of $2.5\cdot10^5$~yr and a $\upsilon_{\mathrm{trans}}\sim240~\mathrm{km~s^{-1}}$ (assuming a 10~kpc distance) or a $\upsilon_{\mathrm{trans}}\sim75~\mathrm{km~s^{-1}}$ (assuming a 3.12~kpc distance). The J1358-6025 (4FGL J1358.3-6026) pulsar, which is located well outside of the remnant's extension ($1.88\degree$ from the remnant's center), has an age of $3.2\cdot10^5$~yr and a $\upsilon_{\mathrm{trans}}\sim313~\mathrm{km~s^{-1}}$ (assuming a 3.12~kpc distance). However, assuming a 10~kpc distance, we derived a $\upsilon_{\mathrm{trans}}\sim1000~\mathrm{km~s^{-1}}$ for the latter pulsar, and thus a potential association with the remnant is forbidden if the latter lies at a $10$~kpc distance, and in general at distances above 5~kpc. The latter two pulsars appear as the most prominent candidates to be associated with the SNR G309.8+00.0, only if the remnant is not of supernova (SN) type Ia progenitor origin. Based on the ages of the nearby pulsars, a distance of 10~kpc appears reasonable for this SNR.
\section{$\gamma$-ray \textit{Fermi}-LAT data analysis}
\label{sec:intro3}

\subsection{$\gamma$-ray observation and data reduction}\label{gammared}

We selected $\sim15.5$~yr from 239557417 to 731081752 mission elapsed time at the start and end of the observation -- August 4, 2008 (15:43:36.0) to March 2, 2024 (14:15:46.0), start and end coordinated universal time (UTC) of the observation -- of Pass 8 (P8R3) 'SOURCE' class data, front and back interactions included (evclass=128, evtype=3), collected with \textit{Fermi}-LAT to inspect the emission at and around the SNR G309.8+00.0 and examine a potential association with the unidentified 4FGL J1349.5-6206c $\gamma$-ray source. The Fermitools\footnote{\url{https://fermi.gsfc.nasa.gov/ssc/data/analysis/scitools/references.html}} version 2.0.8 was employed for data reduction and analysis purposes. A region of interest (ROI) of $20\degree\times20\degree$ centered at the coordinates of the 4FGL J1349.5-6206c $\gamma$-ray source (RA: $207.4\degree$, Dec: $-62.1\degree$) was selected. The latter ROI was selected since it encompassed both the extension of the unidentified 4FGL J1349.5-6206c source and the SNR G309.8+00.0. For the imaging analysis, we restricted ourselves above 1~GeV due to the limited spatial resolution of the instrument at lower energies, given the small angular size of the SNR. This way we ensured that the $\gamma$-ray emission from the SNR G309.8+00.0, of $0\degree.43\times0\degree.32$ size, can be resolved given the improved point spread function (PSF) of the instrument at higher energies. A broader 100 MeV-100 GeV energy range is adopted for the spectral analysis. To remove Earth limb emission which contaminates our signal, we set an upper cut and exclude events with zenith angles above $90\degree$. The standard binned likelihood analysis of Fermitools was used with an $0.025\degree$ bin size to secure a good sampling of the \textit{Fermi}-LAT PSF. The background modeling is achieved by using the Galactic diffuse component $\texttt{(gll\_iem\_v07.fits)}$ and the isotropic diffuse component $\texttt{(iso\_P8R3\_SOURCE\_V3\_v1.txt)}$ as background sources in our model. In addition, all sources included in the recently released incremental 4FGL-DR4 \citep{2023arXiv230712546B,2022ApJS..260...53A} \textit{Fermi}-LAT 14-year source catalog and contained within our ROI were included in the model. Spectral parameters of sources within $5\degree$ from the ROI's center were let vary along with the normalizations of the Galactic and isotropic diffuse components. The remaining parameters were fixed to the catalog values. For imaging analysis purposes above 5~GeV and 10~GeV, only the normalization of the sources and the normalizations of the two background components within a $2\degree$ region from the ROI's center were allowed to vary due to the improved PSF performance\footnote{refer to the SLAC page \url{https://www.slac.stanford.edu/exp/glast/groups/canda/lat_Performance.htm} for further details on the analysis performance of the instrument}.

\begin{figure*}[h!]
    \centering
    
    \includegraphics[width=0.497\textwidth,clip=true, trim= 1.27cm 0.1cm 0.9cm 1.7cm]{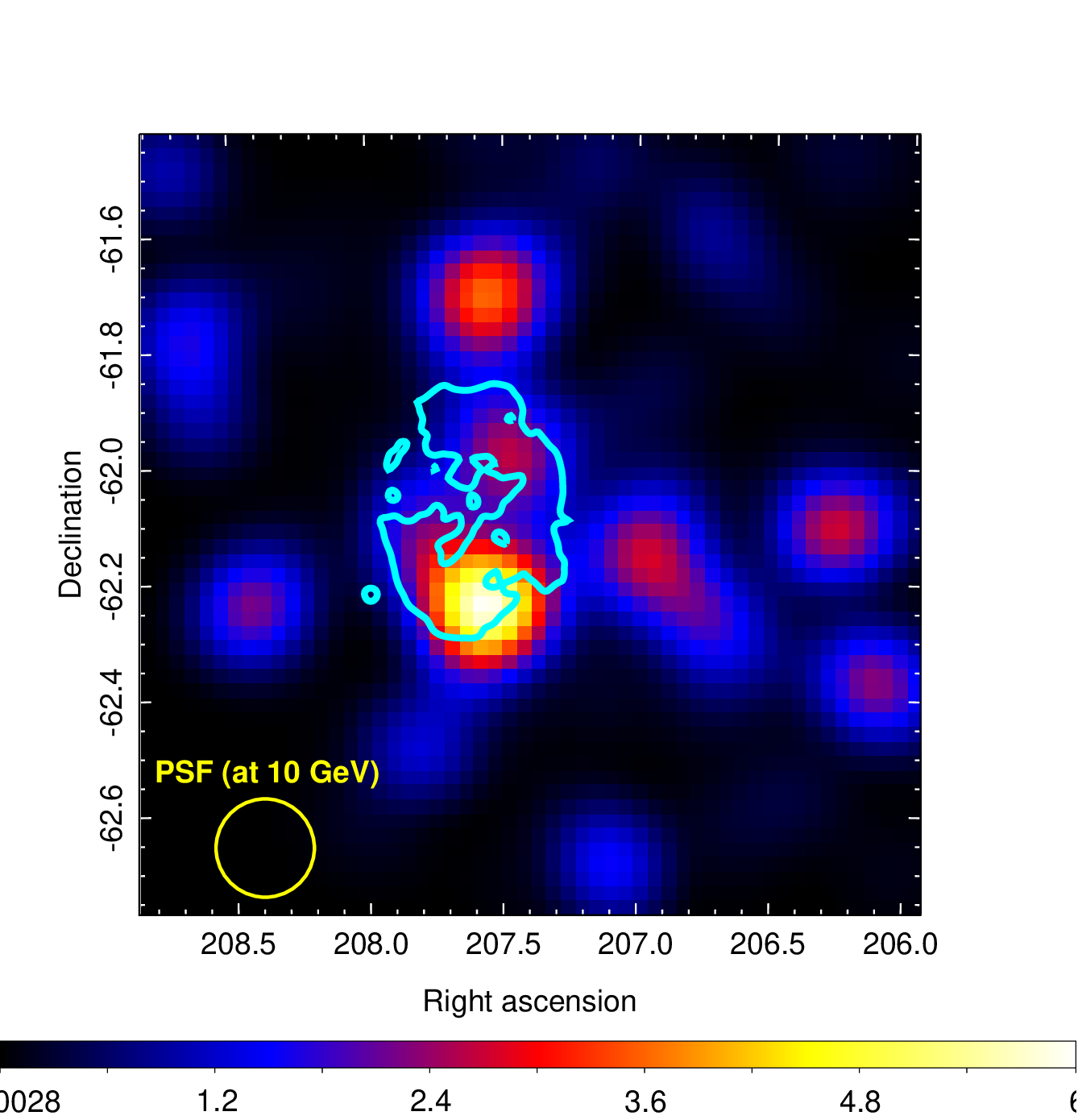}
    \includegraphics[width=0.497\textwidth,clip=true, trim= 1.26cm 0.1cm 0.9cm 1.7cm]{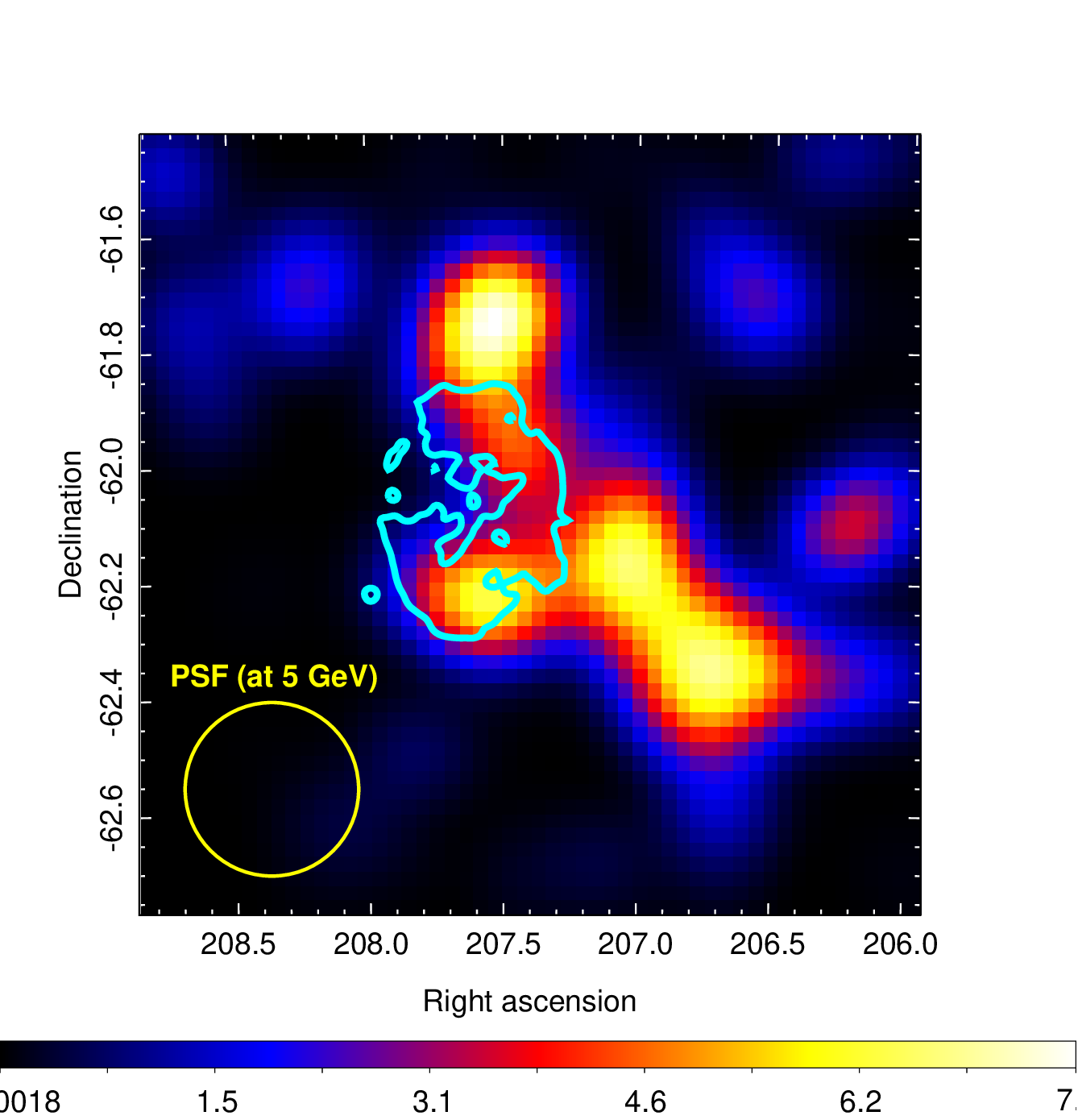}
    \includegraphics[width=0.497\textwidth,clip=true, trim= 1.2cm 0.1cm 0.9cm 1.7cm]{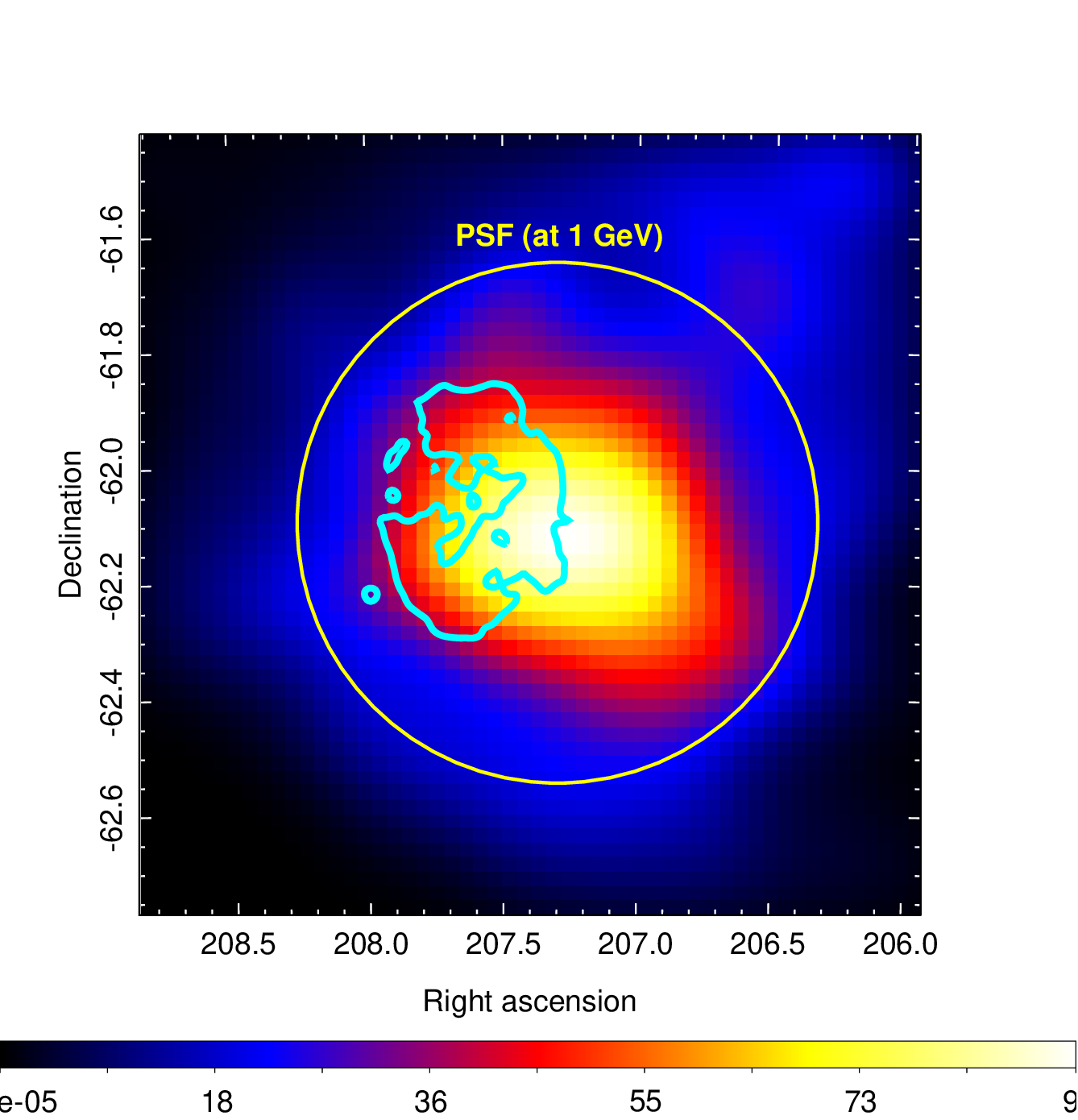}
    \includegraphics[width=0.4975\textwidth,clip=true, trim= 0.9cm 0.1cm 1.0cm 0.8cm]{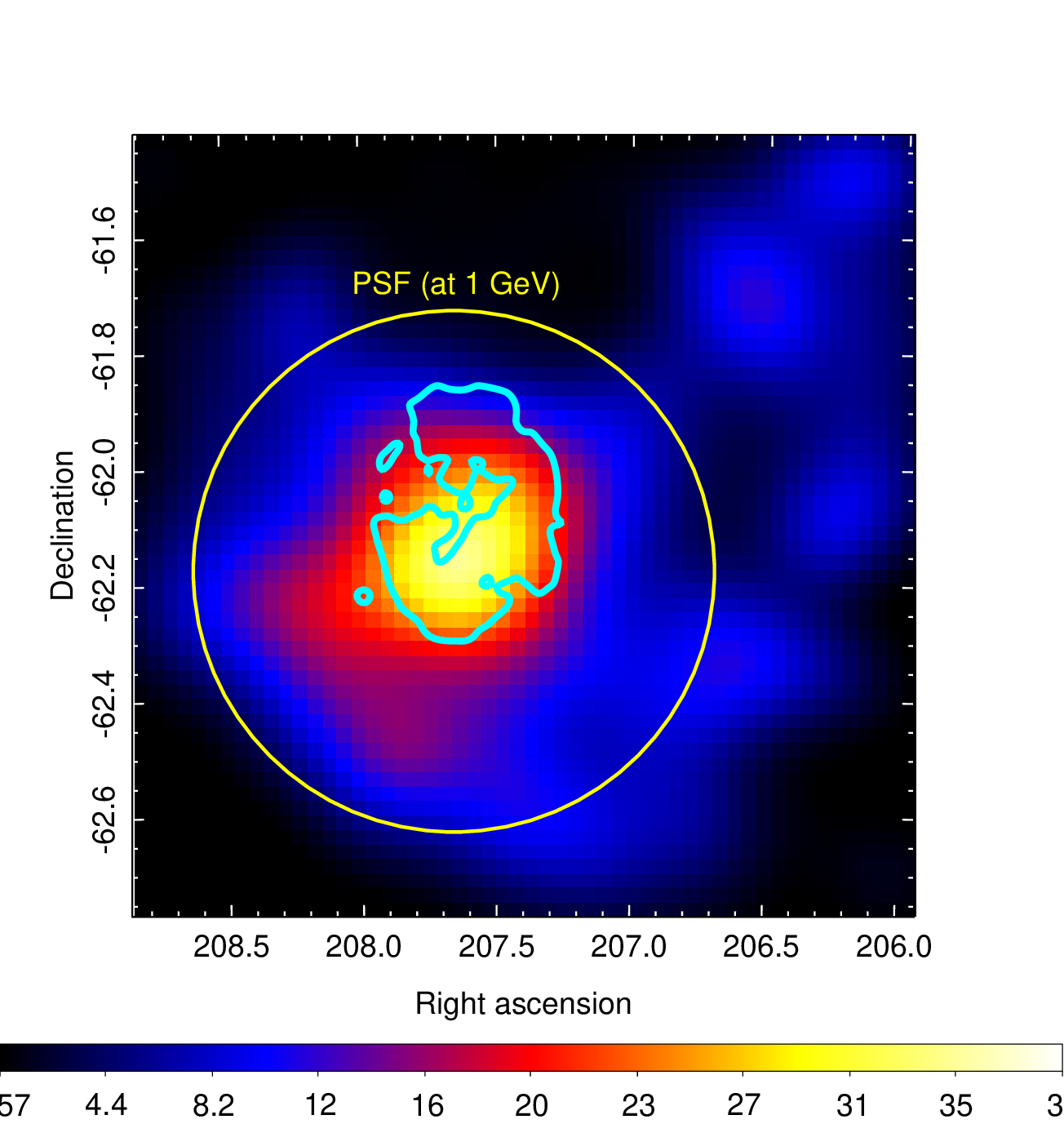}
    \caption{$1\degree.33\times1\degree.33$ \textit{Fermi}-LAT TS maps from the location of the SNR G309.8+00.0 at different energy ranges. All sky maps, of $90''$ pixel size, centered at the best-fit coordinates of the 4FGL J1349.5-6206c source have been smoothed with a 2.5$\sigma$ Gaussian kernel. Upper left panel:  \textit{Fermi}-LAT TS map above 10~GeV. Upper right panel: \textit{Fermi}-LAT TS map above 5~GeV.  Lower left panel:  \textit{Fermi}-LAT TS map above 1~GeV. Lower right panel:  \textit{Fermi}-LAT TS map above 1~GeV when modeling the src-north, src-west, and src-northwest components of the 4FGL J1349.5-6206c source. 
    The cyan contour marks the SNR G309.8+00.0 extension as seen in SUMSS 843~MHz radio data. The yellow circle represents the $68\%$ containment size of the PSF at the energy threshold of each sky map, with no smoothing applied.} 
    \label{Fermi}
\end{figure*}

\subsection{$\gamma$-ray spatial morphology and SED}\label{gammased}

Aiming at examining the $\gamma$-ray emission from the 4FGL J1349.5-6206c -- designation c stands for a source found in a region with bright and/or possibly incorrectly modeled diffuse emission -- and any other sources in our ROI that were missed and were not cataloged in 4FGL-DR4, we created both residuals and Test Statistic (TS) maps at three distinct energy ranges: 1~GeV-1~TeV, 5~GeV-1~TeV, and 10~GeV-1~TeV. Both the residual count maps and the TS maps in the latter energy ranges were produced by modeling the emission from all 4FGL-DR4 sources and the two background components within our ROI. The residual count maps were inspected to ensure that our source(s) of interest, which is(are) located at the center of our ROI, appear(s) to be the brightest and there are no strong negative residuals from the subtracted sources. In Fig.~\ref{Fermi} we present the obtained TS maps in the three different energy ranges. On each panel, the PSF size at the lower energy cut is provided to demonstrate the resolution capacity of the instrument in each sky map. We cut each obtained TS map to a $1\degree.33\times1\degree.33$ size to focus on the emission at and around the small angular size SNR G309.8+00.0.
\begin{figure*}
    \centering
    
    \includegraphics[width=0.33\textwidth,clip=true, trim= 0.75cm 0.1cm 0.9cm 1.7cm]{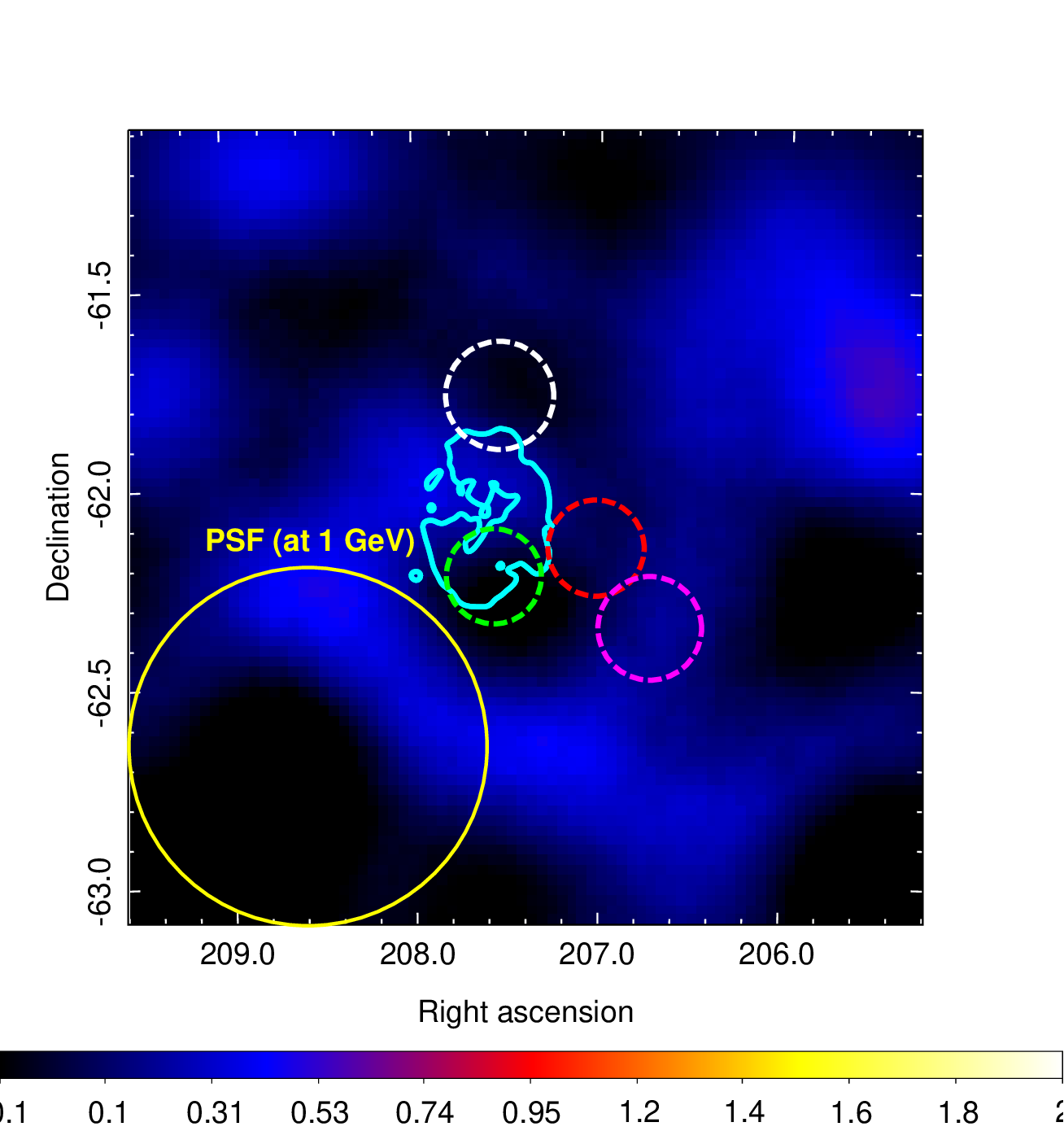}
    \includegraphics[width=0.33\textwidth,clip=true, trim= 0.75cm 0.1cm 0.9cm 1.7cm]{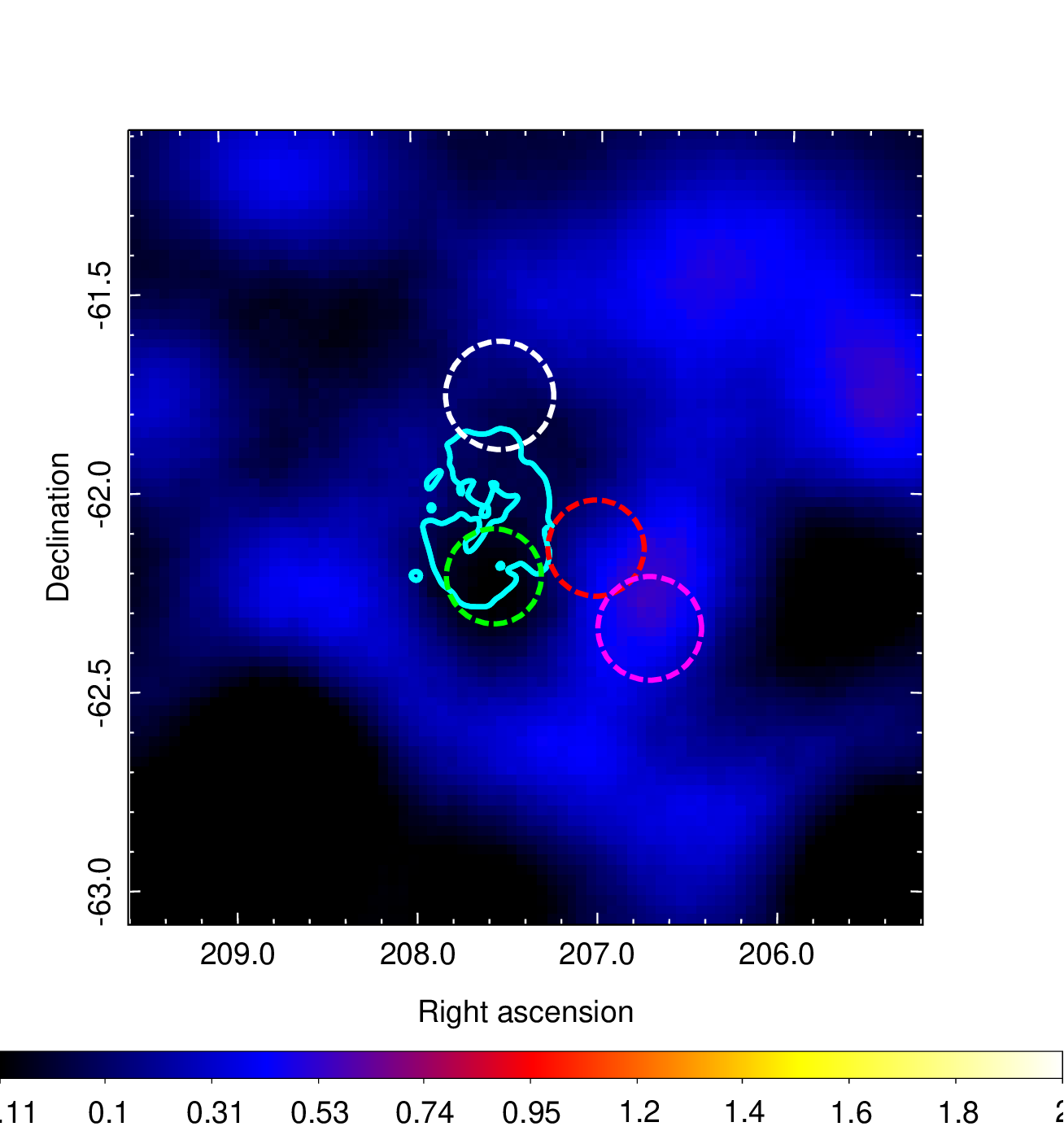}
    \includegraphics[width=0.33\textwidth,clip=true, trim= 0.75cm 0.1cm 0.9cm 1.7cm]{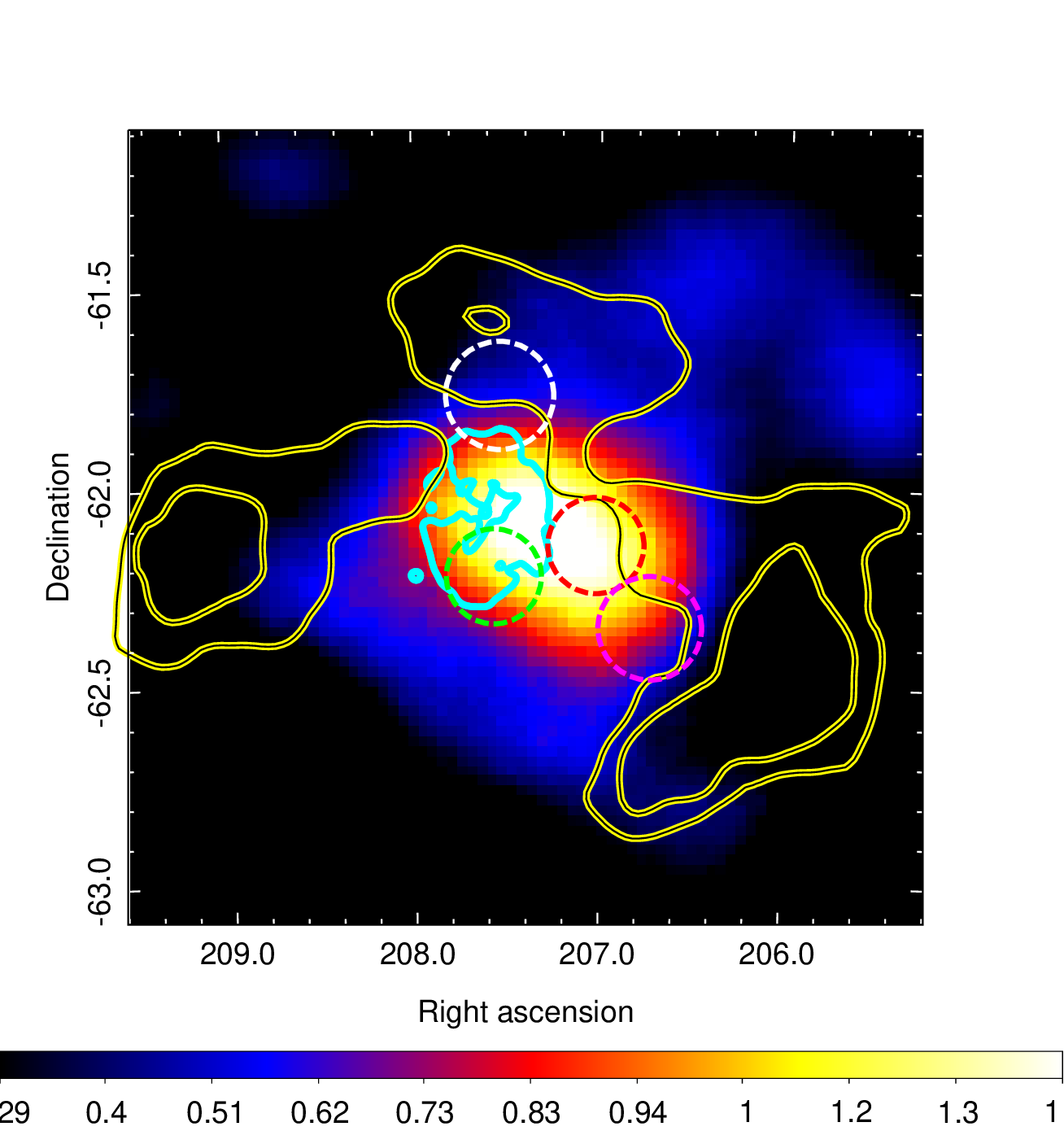}
    \caption{$2\degree.0\times2\degree.0$ \textit{Fermi}-LAT residual count maps above 1~GeV (in units of counts per pixel), from the location of the SNR G309.8+00.0, applying different modeling to the $\gamma$-ray emission from the 4FGL J1349.5-6206c source. All sky maps, of $90''$ pixel size, centered at the best-fit coordinates of the 4FGL J1349.5-6206c source have been smoothed with a 6.5$\sigma$ Gaussian kernel. Left panel:  residual map obtained by modeling the 4FGL J1349.5-6206c as four point-like sources. Middle panel: residual map obtained by modeling the 4FGL J1349.5-6206c as a single source according to the 4FGL-DR4 catalog model. Right panel: residual map obtained by excluding from the model the 4FGL J1349.5-6206c source. The yellow and black contours mark the position of nearby CO clouds as obtained from the CO Galactic plane map \citep{2001ApJ...547..792D}. 
    In all panels, the cyan contour marks the SNR G309.8+00.0 extension as seen in SUMSS 843~MHz radio data. The green, white, red, and magenta dashed circles (centered on the maximum TS values and with radius obtained setting a lower TS cut of $2\sigma$, $>5$~GeV) provide the locations of the 4 components (src, src-north, src-west, and src-southwest, respectively) that the 4FGL J1349.5-6206c source can be decomposed to. The yellow circle represents the $68\%$ containment size of the PSF at the energy threshold of each sky map, with no smoothing applied. The latter PSF size applies to all three panels, however, since it appears that the instrument's PSF cannot resolve individual components of the 4FGL J1349.5-6206c source at $\sim1$~GeV, we only present it on the left-hand panel to avoid overcrowding.} 
    \label{Fermi1}
\end{figure*}
\begin{figure}[h]
    \centering
    \includegraphics[width=0.5\textwidth]{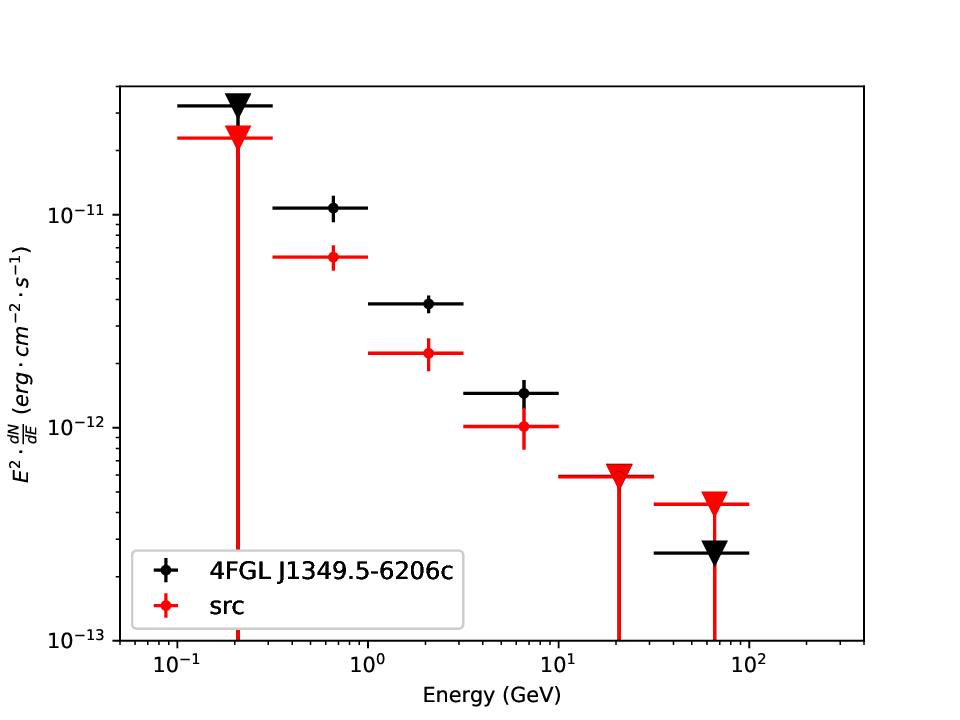}
    \caption{4FGL J1349.5-6206c and src \textit{Fermi}-LAT SED in the 0.1-100~GeV energy range. Black dots correspond to the \textit{Fermi}-LAT spectrum of the 4FGL J1349.5-6206c $\gamma$-ray source. Red dots represent the \textit{Fermi}-LAT spectrum of the src (spatially coincident source to the SNR G309.8+00.0) $\gamma$-ray source. As expected from the obtained TS maps of Fig.~\ref{Fermi}, the $\gamma$-ray emission of the 4FGL J1349.5-6206c $\gamma$-ray source is dominated by the src component at higher energies.}
    \label{Fermi2}
\end{figure}

We started examining the spatial morphology of the $\gamma$-ray emission by inspecting the "richest" data sets $>1$~GeV. As shown in Fig.~\ref{Fermi}, there is a single centroid, detected with a 9.8$\sigma$ significance, spatially coincident with the SNR's location. However, due to the size of the PSF at those energies, we cannot conclude if the latter gigaelectronvolt source, named 4FGL J1349.5-6206c, is a single source or if the $\gamma$-ray emission is a conglomerate of several subsources. Since there is no strong observational evidence that supports the presence of dense material at the western side of the SNR (as seen on the right panel of Fig.~\ref{Fermi1} and discussed at the end of this section) that could subsequently support a hadronically induced $\gamma$-ray scenario, where accelerated nuclei interact with atomic nuclei originating from the dense material giving birth to $\gamma$-ray emission through $\pi^{0}$-decay, and due to the puzzling nature of $\gamma$-ray emission from such an evolved SNR; in this work, we also exploited the alternative scenario that the 4FGL J1349.5-6206c source consists of distinct source components. Therefore, as a next step, we inspected the emission above 5~GeV. The improved PSF size at those energies allowed us to identify at least three distinct source components from this region since there are three centroids whose angular separation exceeds the instrument's PSF size. In particular, we detected the src-north component (RA: $207.53\degree$, Dec:$-61.75\degree$) with 3$\sigma$ significance, the src-southwest component (RA: $207.75\degree$, Dec: $-62.35\degree$) with a 2.7$\sigma$ significance, and the src component (RA: $207.56\degree$, Dec: $-62.23\degree$) with a $2.5\sigma$ significance. Additionally, a potential fourth component, the src-west (RA: $207.04\degree$, Dec: $-62.16\degree$) is detected with a $2.7\sigma$ significance. With src we denote the component that spatially coincides with the SNR G309.8+00.0. Despite the reduced photon statistics, we also inspected $\gamma$-ray emission from the location of interest above 10~GeV which confirms the presence of four distinct sources. The src-west component is now clearly resolved as a separate centroid due to the improved PSF. The src-southwest component is no longer detectable above 10~GeV. Among the latter components, the only significant one $>10$~GeV is the src detected with a $2.5\sigma$ significance. We note that the location of the src component coincides with the brightest part of the remnant's X-ray emitting shell.

For each subcomponent of the 4FGL J1349.5-6206c source, a point-like source spatial template yields a likelihood quality fit of the same goodness as when modeled as extended sources. Thus, we replaced the 4FGL J1349.5-6206c source in our model with four distinct point-like sources and reran the likelihood fit. This way we obtained the residual count map above 1~GeV which perfectly models the emission of 4FGL J1349.5-6206c, as shown in Fig.~\ref{Fermi1}, and also the TS map above 1~GeV having modeled the 4FGL J1349.5-6206c subcomponents that are not spatially coincident with the SNR G309.8+00.0, as shown on the lower right panel of Fig.~\ref{Fermi}. The src component is detected $>1$~GeV with a 5.8~$\sigma$ significance. When treating the src as extended and the rest three as point-like sources no significant improvement in the fit quality is obtained. 

The SED of the src object adopting a LogParabola spectral model is presented in Fig.~\ref{Fermi2} along with the SED of the 4FGL J1349.5-6206c $\gamma$-ray source (when treated as a single source) adopting the spatial and spectral model provided in the 4FGL-DR4 catalog (spectral parameters: $\alpha$=3.0, $\beta$=0.2). A powerlaw model describes well the gigaelectronvolt SED of both sources, however, a LogParabola model is preferred over a simple powerlaw based on the \textit{Signif\_Curve} Fermitools task. The shapes and the narrow energy range of the gigaelectronvolt SED of both sources do not allow us to discriminate between a hadronic and a leptonic mechanism for the production of $\gamma$ rays. However, the remnant's evolved stage favors a hadronic interpretation. If future studies confirm an age as large as $10^5$~yr, the nature of the $\gamma$-ray emission becomes even more puzzling, placing this remnant among the oldest $\gamma$-ray loud SNRs ever detected, along with, for example, the G279.0+01.1 \citep{2024A&A...685A..23M} and the S147 \citep{2024arXiv240117312M}. Particle acceleration at such large ages is questionable due to the rapid decrease of the shock speed at the latter stages of the SNRs evolution. However, recent theoretical works, for example, \citet{2019ApJ...876...27Y}, demonstrate that under specific conditions, usually low-density environments (e.g., SNR-in-cavity scenario, see \citet{2024arXiv240117261K} for a showcase example), even leptonically induced $\gamma$-ray emission could be observed from old SNRs.
Since there is still uncertainty in terms of which parts of the emission of the 4FGL J1349.5-6206c source are associated with the SNR G309.8+00.0 in combination with the absence of X-ray emission of nonthermal nature (X-ray synchrotron emission), we do not provide a multiwavelength SED of the latter source. This is a task for the future.

We also examined our ROI for CO clouds that could potentially interact with accelerated nuclei originating from the SNR G309.8+00.0 and thus yielding $\gamma$-ray emission. No CO clouds were found to be spatially coincident with the 4FGL J1349.5-6206c $\gamma$-ray source as shown on the right panel of Fig.~\ref{Fermi1}. In addition, the inspection of AKARI and the Wide-field Infrared Survey Explorer (WISE) infrared (IR) data from the location of the remnant did not yield the detection of enhanced IR emission regions to the western part of the SNR, and thus, there is no evidence that would make us believe that the $\gamma$-ray emission stems from the interaction of the shocks with material.

\section{Discussion and conclusions}
\label{sec:intro4}

We present the first study of the well-established, from radio observations, SNR G309.8+00.0 at higher energies. We report the detection of its X-ray counterpart using data from the first four eROSITA all-sky surveys (eRASS:4) and investigate the nature of the unidentified 4FGL J1349.5-6206c $\gamma$-ray source using 15.5~yr of \textit{Fermi}-LAT data. We conclude that the SNR is most likely to account for at least a significant portion (if not all) of the emission from the latter $\gamma$-ray source. This work is evident of the importance of the multiwavelength study of supernova remnants in determining key properties of their nature.

Utilizing eRASS:4 data we provide the first X-ray view of the SNR 309.8+00.0. The X-ray emission is mainly confined to the 1-2~keV energy range. The spatial morphology of the X-ray emission is best fitted by an ellipse of $0\degree.43\times0\degree.32$ in size and its shell-type appearance is in good spatial correlation with the radio synchrotron emission from the SNR. The X-ray emission mainly fills the southern half of the remnant's shell. The SNR exhibits limb-brightening features since it appears brighter at the edges of the shell compared to when looking through its central parts. Moving to the physical processes and the physics interpretation behind the above imaging results, we conclude that the nature of the X-ray emission originating from the SNR is purely thermal and likely in equilibrium. However, nonequilibrium models provide spectral fits of equal goodness and thus cannot be excluded due to the limited statistics of the X-ray data. Either there is an absence of a continuum nonthermal component or the faint emission of the SNR does not allow its detection with an all-sky survey of limited exposure such as eROSITA. The remnant's X-ray spectrum is well fitted by the tbabs$\times$vapec thermal plasma model in equilibrium with a kT=$0.34\pm0.1$~keV and a
$N_{\mathrm{H}}=3.1_{-0.5}^{+0.7}~10^{22}~\mathrm{cm^{-2}}$. Despite the limited statistics, we detect relatively prominent Mg XI+XII (1.3-1.5~keV) and Si XIII (1.7-1.9~keV) emission lines across the remnant's area. The silicon emission is mainly confined to the southeastern portion of the remnant whereas the magnesium emission is more uniformly distributed on the ring-like structure of the X-ray shell. 

Due to the strong absorption at soft X-rays ($<1$~keV), no detailed comparison of the elemental abundance fractions can be made. Thus, we cannot disentangle between a thermonuclear and a core-collapse (CC) progenitor origin. 
Even though the shell-type morphology of typical Ia progenitor origin SNRs is usually highly symmetrical mainly due to the lack of a massive progenitor star, that exhibits strong stellar winds that disturb the surrounding medium, this does not come as a requirement (e.g., refer to the Kepler SNR \citep{2012A&A...537A.139C} and the SNR G321.3-3.9 \citep{Mantovanini}). In contrast to the SNR G309.8+00.0, typical type Ia SNRs have been mainly found in less dense medium away from the Galactic plane which would explain their symmetrical shape mainly due to the lower density medium within which they expand. However, there have been reports of type Ia SNRs with disturbed shapes that are located close to the Galactic plane. The SNR G321.3-3.9 is a showcase example of an elliptical-shaped SNR that lies close to the Galactic plane and is believed to be of type Ia progenitor origin, as reported by \citet{Mantovanini}. Therefore, the location of the SNR G309.8+00.0 would explain its elliptical, and thus less symmetrical, shape. A deep X-ray observation of the remnant is necessary to investigate in detail its X-ray spectrum and consequently its progenitor nature. In addition, optical studies of the remnant would be essential to examine the presence of radiative shocks in dense knots that would be evident of circumstellar material (CSM) shed by the progenitor. Toward this end, we also investigated the full-sky H$\alpha$ map with 6' (FWHM) resolution \citep{2003ApJS..146..407F} from the remnant's location, but no optical counterpart was found.

We also provide a distance and age estimate of the SNR using the absorption column density values derived from its X-ray spectral fit. The distance value on the order of 10~kpc that we derived, by exploiting the latest optical extinction data sets by \citet{2022A&A...661A.147L} and the derived $N_\mathrm{H}$ values, appears to be inconsistent with previous distance measurements of the remnant (3.12~kpc \citep{2020A&A...639A..72W}). The large uncertainties in the methods employed, both in this work and in the literature, for the distance computation force us to discuss age estimates when considering both a 3.12~kpc and a 10~kpc distance. We obtain an age on the order of $10^4$~yr and an age on the order of $10^5$~yr in the two distinct cases, respectively. 

In the case of a CC-type SN progenitor, we also examined potential associations with nearby pulsars. The J1350-6225 (4FGL J1350.6-6224) pulsar, which is located just $0.35\degree$ away from the remnant's center, and the J1358-6025 (4FGL J1358.3-6026) pulsar located $1.88\degree$ away from the remnant's center (both outside the remnant's extension) appear to be the most probable candidates based on their age and transverse velocity estimates that would be required to reach their present location assuming that they started traveling from the center of the remnant.

Employing 15.5~years of \textit{Fermi}-LAT data, and carrying a detailed spatial data analysis of the $\gamma$-ray emission at and around the SNR G309.8+00.0, we show that the emission from the unidentified 4FGL J1349.5-6206c source can either be modeled as a single source that is likely associated with the SNR or be decomposed to four point-like components. In the latter case, the src component which is spatially coincident with the southern part of the SNR G309.8+00.0 (which is also the brightest -- enhanced X-ray emission -- part of the remnant's X-ray emitting shell) is detected with a 5.8$\sigma$ significance above 1~GeV. Thus, we conclude that at least a significant part of the $\gamma$-ray emission (if not all), named after 4FGL J1349.5-6206c, is likely associated with the SNR G309.8+00.0. The spectral shape of both the src component and the entire 4FGL J1349.5-6206c source appears to be best fitted with a LogParabola. However, a powerlaw model cannot be ruled out. The spatial component of the src component can be well fitted with both a point-like and an extended morphology. An improved PSF size is required to inspect in detail the emission spatial components $<5$~GeV aiming at addressing the bias of the limited photon statistics at higher energies and consequently disentangle between a single source or multiple distinct source components. The shape and narrow energy range to which both the src component and the 4FGL J1349.5-6206c source gigaelectronvolt SED extends prevent us from providing a more complete description of the nature of the $\gamma$-ray emission and distinguishing between hadronic and leptonic origin of the $\gamma$-ray photons. However, the relatively old remnant's age of $10^4-10^5$~yr supports a hadronic $\gamma$-ray interpretation.



This analysis demonstrates the importance of multiwavelength studies in determining the key properties of Galactic SNRs. An in-depth examination of the SNR G309.8+00.0 is, however, prohibited due to its location. To fully understand the true physical processes behind the emission mechanics of this SNR, future studies utilizing longer exposure times in X-rays and an improved instrument PSF at $\sim\mathrm{gigaelectronvolt}$ energies will be required.

\noindent\textit{Acknowledgements}

M.M. and G.P. acknowledge support from the Deutsche Forschungsgemeinschaft through grant PU 308/2-1.

This work is based on data from eROSITA, the soft X-ray instrument aboard SRG, a joint Russian-German science mission supported by the Russian Space Agency (Roskosmos), in the interests of the Russian Academy of Sciences represented by its Space Research Institute (IKI), and the Deutsches Zentrum für Luft- und Raumfahrt (DLR). The SRG spacecraft was built by Lavochkin Association (NPOL) and its subcontractors, and is operated by NPOL with support from the Max Planck Institute for Extraterrestrial Physics (MPE).

The development and construction of the eROSITA X-ray instrument was led by MPE, with contributions from the Dr. Karl Remeis Observatory Bamberg $\&$ ECAP (FAU Erlangen-Nuernberg), the University of Hamburg Observatory, the Leibniz Institute for Astrophysics Potsdam (AIP), and the Institute for Astronomy and Astrophysics of the University of Tübingen, with the support of DLR and the Max Planck Society. The Argelander Institute for Astronomy of the University of Bonn and the Ludwig Maximilians Universität Munich also participated in the science preparation for eROSITA.
The eROSITA data shown here were processed using the eSASS/NRTA software system developed by the German eROSITA consortium.

We thank the EXPLORE team which provided us access to the G-TOMO tool of the EXPLORE platform \url{https://explore-platform.eu/} allowing us to exploit updated GAIA-2MASS data.


\bibliography{biblio}

\begin{thebibliography}{45}
\expandafter\ifx\csname natexlab\endcsname\relax\def\natexlab#1{#1}\fi

\bibitem[{{Abdo} {et~al.}(2010){Abdo}, {Ackermann}, {Ajello}, {Allafort},
  {Antolini}, {Atwood}, {Axelsson}, {Baldini}, {Ballet}, {Barbiellini},
  {Bastieri}, {Baughman}, {Bechtol}, {Bellazzini}, {Belli}, {Berenji},
  {Bisello}, {Blandford}, {Bloom}, {Bonamente}, {Bonnell}, {Borgland},
  {Bouvier}, {Bregeon}, {Brez}, {Brigida}, {Bruel}, {Burnett}, {Busetto},
  {Buson}, {Caliandro}, {Cameron}, {Campana}, {Canadas}, {Caraveo}, {Carrigan},
  {Casandjian}, {Cavazzuti}, {Ceccanti}, {Cecchi}, {{\c{C}}elik}, {Charles},
  {Chekhtman}, {Cheung}, {Chiang}, {Cillis}, {Ciprini}, {Claus},
  {Cohen-Tanugi}, {Conrad}, {Corbet}, {Davis}, {DeKlotz}, {den Hartog},
  {Dermer}, {de Angelis}, {de Luca}, {de Palma}, {Digel}, {Dormody}, {Silva},
  {Drell}, {Dubois}, {Dumora}, {Fabiani}, {Farnier}, {Favuzzi}, {Fegan},
  {Ferrara}, {Focke}, {Fortin}, {Frailis}, {Fukazawa}, {Funk}, {Fusco},
  {Gargano}, {Gasparrini}, {Gehrels}, {Germani}, {Giavitto}, {Giebels},
  {Giglietto}, {Giommi}, {Giordano}, {Giroletti}, {Glanzman}, {Godfrey},
  {Grenier}, {Grondin}, {Grove}, {Guillemot}, {Guiriec}, {Gustafsson},
  {Hadasch}, {Hanabata}, {Harding}, {Hayashida}, {Hays}, {Healey}, {Hill},
  {Horan}, {Hughes}, {Iafrate}, {J{\'o}hannesson}, {Johnson}, {Johnson},
  {Johnson}, {Johnson}, {Kamae}, {Katagiri}, {Kataoka}, {Kawai}, {Kerr},
  {Kn{\"o}dlseder}, {Kocevski}, {Kuss}, {Lande}, {Landriu}, {Latronico}, {Lee},
  {Lemoine-Goumard}, {Lionetto}, {Llena Garde}, {Longo}, {Loparco}, {Lott},
  {Lovellette}, {Lubrano}, {Madejski}, {Makeev}, {Marangelli}, {Marelli},
  {Massaro}, {Mazziotta}, {McConville}, {McEnery}, {Michelson}, {Minuti},
  {Mitthumsiri}, {Mizuno}, {Moiseev}, {Mongelli}, {Monte}, {Monzani},
  {Moretti}, {Morselli}, {Moskalenko}, {Murgia}, {Nakajima}, {Nakamori},
  {Naumann-Godo}, {Nolan}, {Norris}, {Nuss}, {Ohno}, {Ohsugi}, {Omodei},
  {Orlando}, {Ormes}, {Ozaki}, {Paccagnella}, {Paneque}, {Panetta}, {Parent},
  {Pelassa}, {Pepe}, {Pesce-Rollins}, {Pinchera}, {Piron}, {Porter}, {Poupard},
  {Rain{\`o}}, {Rando}, {Ray}, {Razzano}, {Razzaque}, {Rea}, {Reimer},
  {Reimer}, {Reposeur}, {Ripken}, {Ritz}, {Rochester}, {Rodriguez}, {Romani},
  {Roth}, {Sadrozinski}, {Salvetti}, {Sanchez}, {Sander}, {Saz Parkinson},
  {Scargle}, {Schalk}, {Scolieri}, {Sgr{\`o}}, {Shaw}, {Siskind}, {Smith},
  {Smith}, {Spandre}, {Spinelli}, {Starck}, {Stephens}, {Striani}, {Strickman},
  {Strong}, {Suson}, {Tajima}, {Takahashi}, {Takahashi}, {Tanaka}, {Thayer},
  {Thayer}, {Thompson}, {Tibaldo}, {Tibolla}, {Tinebra}, {Torres}, {Tosti},
  {Tramacere}, {Uchiyama}, {Usher}, {Van Etten}, {Vasileiou}, {Vilchez},
  {Vitale}, {Waite}, {Wallace}, {Wang}, {Watters}, {Winer}, {Wood}, {Yang},
  {Ylinen}, {Ziegler}, \& {Fermi LAT Collaboration}}]{2010ApJS..188..405A}
{Abdo}, A.~A., {Ackermann}, M., {Ajello}, M., {et~al.} 2010, \apjs, 188, 405

\bibitem[{{Abdollahi} {et~al.}(2022){Abdollahi}, {Acero}, {Baldini}, {Ballet},
  {Bastieri}, {Bellazzini}, {Berenji}, {Berretta}, {Bissaldi}, {Blandford},
  {Bloom}, {Bonino}, {Brill}, {Britto}, {Bruel}, {Burnett}, {Buson}, {Cameron},
  {Caputo}, {Caraveo}, {Castro}, {Chaty}, {Cheung}, {Chiaro}, {Cibrario},
  {Ciprini}, {Coronado-Bl{\'a}zquez}, {Crnogorcevic}, {Cutini}, {D'Ammando},
  {De Gaetano}, {Digel}, {Di Lalla}, {Dirirsa}, {Di Venere}, {Dom{\'\i}nguez},
  {Fallah Ramazani}, {Fegan}, {Ferrara}, {Fiori}, {Fleischhack}, {Franckowiak},
  {Fukazawa}, {Funk}, {Fusco}, {Galanti}, {Gammaldi}, {Gargano}, {Garrappa},
  {Gasparrini}, {Giacchino}, {Giglietto}, {Giordano}, {Giroletti}, {Glanzman},
  {Green}, {Grenier}, {Grondin}, {Guillemot}, {Guiriec}, {Gustafsson},
  {Harding}, {Hays}, {Hewitt}, {Horan}, {Hou}, {J{\'o}hannesson}, {Karwin},
  {Kayanoki}, {Kerr}, {Kuss}, {Landriu}, {Larsson}, {Latronico},
  {Lemoine-Goumard}, {Li}, {Liodakis}, {Longo}, {Loparco}, {Lott}, {Lubrano},
  {Maldera}, {Malyshev}, {Manfreda}, {Mart{\'\i}-Devesa}, {Mazziotta}, {Mereu},
  {Meyer}, {Michelson}, {Mirabal}, {Mitthumsiri}, {Mizuno}, {Moiseev},
  {Monzani}, {Morselli}, {Moskalenko}, {Negro}, {Nuss}, {Omodei}, {Orienti},
  {Orlando}, {Paneque}, {Pei}, {Perkins}, {Persic}, {Pesce-Rollins},
  {Petrosian}, {Pillera}, {Poon}, {Porter}, {Principe}, {Rain{\`o}}, {Rando},
  {Rani}, {Razzano}, {Razzaque}, {Reimer}, {Reimer}, {Reposeur},
  {S{\'a}nchez-Conde}, {Saz Parkinson}, {Scotton}, {Serini}, {Sgr{\`o}},
  {Siskind}, {Smith}, {Spandre}, {Spinelli}, {Sueoka}, {Suson}, {Tajima},
  {Tak}, {Thayer}, {Thompson}, {Torres}, {Troja}, {Valverde}, {Wood}, \&
  {Zaharijas}}]{2022ApJS..260...53A}
{Abdollahi}, S., {Acero}, F., {Baldini}, L., {et~al.} 2022, \apjs, 260, 53

\bibitem[{{Acero} {et~al.}(2015){Acero}, {Ackermann}, {Ajello}, {Albert},
  {Atwood}, {Axelsson}, {Baldini}, {Ballet}, {Barbiellini}, {Bastieri},
  {Belfiore}, {Bellazzini}, {Bissaldi}, {Blandford}, {Bloom}, {Bogart},
  {Bonino}, {Bottacini}, {Bregeon}, {Britto}, {Bruel}, {Buehler}, {Burnett},
  {Buson}, {Caliandro}, {Cameron}, {Caputo}, {Caragiulo}, {Caraveo},
  {Casandjian}, {Cavazzuti}, {Charles}, {Chaves}, {Chekhtman}, {Cheung},
  {Chiang}, {Chiaro}, {Ciprini}, {Claus}, {Cohen-Tanugi}, {Cominsky}, {Conrad},
  {Cutini}, {D'Ammando}, {de Angelis}, {DeKlotz}, {de Palma}, {Desiante},
  {Digel}, {Di Venere}, {Drell}, {Dubois}, {Dumora}, {Favuzzi}, {Fegan},
  {Ferrara}, {Finke}, {Franckowiak}, {Fukazawa}, {Funk}, {Fusco}, {Gargano},
  {Gasparrini}, {Giebels}, {Giglietto}, {Giommi}, {Giordano}, {Giroletti},
  {Glanzman}, {Godfrey}, {Grenier}, {Grondin}, {Grove}, {Guillemot}, {Guiriec},
  {Hadasch}, {Harding}, {Hays}, {Hewitt}, {Hill}, {Horan}, {Iafrate}, {Jogler},
  {J{\'o}hannesson}, {Johnson}, {Johnson}, {Johnson}, {Johnson}, {Kamae},
  {Kataoka}, {Katsuta}, {Kuss}, {La Mura}, {Landriu}, {Larsson}, {Latronico},
  {Lemoine-Goumard}, {Li}, {Li}, {Longo}, {Loparco}, {Lott}, {Lovellette},
  {Lubrano}, {Madejski}, {Massaro}, {Mayer}, {Mazziotta}, {McEnery},
  {Michelson}, {Mirabal}, {Mizuno}, {Moiseev}, {Mongelli}, {Monzani},
  {Morselli}, {Moskalenko}, {Murgia}, {Nuss}, {Ohno}, {Ohsugi}, {Omodei},
  {Orienti}, {Orlando}, {Ormes}, {Paneque}, {Panetta}, {Perkins},
  {Pesce-Rollins}, {Piron}, {Pivato}, {Porter}, {Racusin}, {Rando}, {Razzano},
  {Razzaque}, {Reimer}, {Reimer}, {Reposeur}, {Rochester}, {Romani},
  {Salvetti}, {S{\'a}nchez-Conde}, {Saz Parkinson}, {Schulz}, {Siskind},
  {Smith}, {Spada}, {Spandre}, {Spinelli}, {Stephens}, {Strong}, {Suson},
  {Takahashi}, {Takahashi}, {Tanaka}, {Thayer}, {Thayer}, {Thompson},
  {Tibaldo}, {Tibolla}, {Torres}, {Torresi}, {Tosti}, {Troja}, {Van Klaveren},
  {Vianello}, {Winer}, {Wood}, {Wood}, {Zimmer}, \& {Fermi-LAT
  Collaboration}}]{2015ApJS..218...23A}
{Acero}, F., {Ackermann}, M., {Ajello}, M., {et~al.} 2015, \apjs, 218, 23

\bibitem[{{Acero} {et~al.}(2016){Acero}, {Ackermann}, {Ajello}, {Baldini},
  {Ballet}, {Barbiellini}, {Bastieri}, {Bellazzini}, {Bissaldi}, {Blandford},
  {Bloom}, {Bonino}, {Bottacini}, {Brandt}, {Bregeon}, {Bruel}, {Buehler},
  {Buson}, {Caliandro}, {Cameron}, {Caputo}, {Caragiulo}, {Caraveo},
  {Casandjian}, {Cavazzuti}, {Cecchi}, {Chekhtman}, {Chiang}, {Chiaro},
  {Ciprini}, {Claus}, {Cohen}, {Cohen-Tanugi}, {Cominsky}, {Condon}, {Conrad},
  {Cutini}, {D'Ammando}, {de Angelis}, {de Palma}, {Desiante}, {Digel}, {Di
  Venere}, {Drell}, {Drlica-Wagner}, {Favuzzi}, {Ferrara}, {Franckowiak},
  {Fukazawa}, {Funk}, {Fusco}, {Gargano}, {Gasparrini}, {Giglietto}, {Giommi},
  {Giordano}, {Giroletti}, {Glanzman}, {Godfrey}, {Gomez-Vargas}, {Grenier},
  {Grondin}, {Guillemot}, {Guiriec}, {Gustafsson}, {Hadasch}, {Harding},
  {Hayashida}, {Hays}, {Hewitt}, {Hill}, {Horan}, {Hou}, {Iafrate}, {Jogler},
  {J{\'o}hannesson}, {Johnson}, {Kamae}, {Katagiri}, {Kataoka}, {Katsuta},
  {Kerr}, {Kn{\"o}dlseder}, {Kocevski}, {Kuss}, {Laffon}, {Lande}, {Larsson},
  {Latronico}, {Lemoine-Goumard}, {Li}, {Li}, {Longo}, {Loparco}, {Lovellette},
  {Lubrano}, {Magill}, {Maldera}, {Marelli}, {Mayer}, {Mazziotta}, {Michelson},
  {Mitthumsiri}, {Mizuno}, {Moiseev}, {Monzani}, {Moretti}, {Morselli},
  {Moskalenko}, {Murgia}, {Nemmen}, {Nuss}, {Ohsugi}, {Omodei}, {Orienti},
  {Orlando}, {Ormes}, {Paneque}, {Perkins}, {Pesce-Rollins}, {Petrosian},
  {Piron}, {Pivato}, {Porter}, {Rain{\`o}}, {Rando}, {Razzano}, {Razzaque},
  {Reimer}, {Reimer}, {Renaud}, {Reposeur}, {Rousseau}, {Saz Parkinson},
  {Schmid}, {Schulz}, {Sgr{\`o}}, {Siskind}, {Spada}, {Spandre}, {Spinelli},
  {Strong}, {Suson}, {Tajima}, {Takahashi}, {Tanaka}, {Thayer}, {Thompson},
  {Tibaldo}, {Tibolla}, {Torres}, {Tosti}, {Troja}, {Uchiyama}, {Vianello},
  {Wells}, {Wood}, {Wood}, {Yassine}, {den Hartog}, \&
  {Zimmer}}]{2016ApJS..224....8A}
{Acero}, F., {Ackermann}, M., {Ajello}, M., {et~al.} 2016, \apjs, 224, 8

\bibitem[{{Ackermann} {et~al.}(2013){Ackermann}, {Ajello}, {Allafort},
  {Baldini}, {Ballet}, {Barbiellini}, {Baring}, {Bastieri}, {Bechtol},
  {Bellazzini}, {Blandford}, {Bloom}, {Bonamente}, {Borgland}, {Bottacini},
  {Brandt}, {Bregeon}, {Brigida}, {Bruel}, {Buehler}, {Busetto}, {Buson},
  {Caliandro}, {Cameron}, {Caraveo}, {Casandjian}, {Cecchi}, {{\c{C}}elik},
  {Charles}, {Chaty}, {Chaves}, {Chekhtman}, {Cheung}, {Chiang}, {Chiaro},
  {Cillis}, {Ciprini}, {Claus}, {Cohen-Tanugi}, {Cominsky}, {Conrad}, {Corbel},
  {Cutini}, {D'Ammando}, {de Angelis}, {de Palma}, {Dermer}, {do Couto e
  Silva}, {Drell}, {Drlica-Wagner}, {Falletti}, {Favuzzi}, {Ferrara},
  {Franckowiak}, {Fukazawa}, {Funk}, {Fusco}, {Gargano}, {Germani},
  {Giglietto}, {Giommi}, {Giordano}, {Giroletti}, {Glanzman}, {Godfrey},
  {Grenier}, {Grondin}, {Grove}, {Guiriec}, {Hadasch}, {Hanabata}, {Harding},
  {Hayashida}, {Hayashi}, {Hays}, {Hewitt}, {Hill}, {Hughes}, {Jackson},
  {Jogler}, {J{\'o}hannesson}, {Johnson}, {Kamae}, {Kataoka}, {Katsuta},
  {Kn{\"o}dlseder}, {Kuss}, {Lande}, {Larsson}, {Latronico}, {Lemoine-Goumard},
  {Longo}, {Loparco}, {Lovellette}, {Lubrano}, {Madejski}, {Massaro}, {Mayer},
  {Mazziotta}, {McEnery}, {Mehault}, {Michelson}, {Mignani}, {Mitthumsiri},
  {Mizuno}, {Moiseev}, {Monzani}, {Morselli}, {Moskalenko}, {Murgia},
  {Nakamori}, {Nemmen}, {Nuss}, {Ohno}, {Ohsugi}, {Omodei}, {Orienti},
  {Orlando}, {Ormes}, {Paneque}, {Perkins}, {Pesce-Rollins}, {Piron}, {Pivato},
  {Rain{\`o}}, {Rando}, {Razzano}, {Razzaque}, {Reimer}, {Reimer}, {Ritz},
  {Romoli}, {S{\'a}nchez-Conde}, {Schulz}, {Sgr{\`o}}, {Simeon}, {Siskind},
  {Smith}, {Spandre}, {Spinelli}, {Stecker}, {Strong}, {Suson}, {Tajima},
  {Takahashi}, {Takahashi}, {Tanaka}, {Thayer}, {Thayer}, {Thompson},
  {Thorsett}, {Tibaldo}, {Tibolla}, {Tinivella}, {Troja}, {Uchiyama}, {Usher},
  {Vandenbroucke}, {Vasileiou}, {Vianello}, {Vitale}, {Waite}, {Werner},
  {Winer}, {Wood}, {Wood}, {Yamazaki}, {Yang}, \&
  {Zimmer}}]{2013Sci...339..807A}
{Ackermann}, M., {Ajello}, M., {Allafort}, A., {et~al.} 2013, Science, 339, 807

\bibitem[{{Ballet} {et~al.}(2023){Ballet}, {Bruel}, {Burnett}, {Lott}, \& {The
  Fermi-LAT collaboration}}]{2023arXiv230712546B}
{Ballet}, J., {Bruel}, P., {Burnett}, T.~H., {Lott}, B., \& {The Fermi-LAT
  collaboration}. 2023, incremental version of the fourth Fermi-LAT catalog
  published in ApJ, arXiv e-prints, arXiv:2307.12546

\bibitem[{Borkowski {et~al.}(2006)Borkowski, Hendrick, \&
  Reynolds}]{Borkowski_2006}
Borkowski, K.~J., Hendrick, S.~P., \& Reynolds, S.~P. 2006, The Astrophysical
  Journal, 652, 1259

\bibitem[{{Borkowski} {et~al.}(2001){Borkowski}, {Lyerly}, \&
  {Reynolds}}]{2001ApJ...548..820B}
{Borkowski}, K.~J., {Lyerly}, W.~J., \& {Reynolds}, S.~P. 2001, \apj, 548, 820

\bibitem[{{Brunner} {et~al.}(2022){Brunner}, {Liu}, {Lamer}, {Georgakakis},
  {Merloni}, {Brusa}, {Bulbul}, {Dennerl}, {Friedrich}, {Liu}, {Maitra},
  {Nandra}, {Ramos-Ceja}, {Sanders}, {Stewart}, {Boller}, {Buchner}, {Clerc},
  {Comparat}, {Dwelly}, {Eckert}, {Finoguenov}, {Freyberg}, {Ghirardini},
  {Gueguen}, {Haberl}, {Kreykenbohm}, {Krumpe}, {Osterhage}, {Pacaud},
  {Predehl}, {Reiprich}, {Robrade}, {Salvato}, {Santangelo}, {Schrabback},
  {Schwope}, \& {Wilms}}]{2022A&A...661A...1B}
{Brunner}, H., {Liu}, T., {Lamer}, G., {et~al.} 2022, \aap, 661, A1

\bibitem[{{Cash}(1979)}]{1979ApJ...228..939C}
{Cash}, W. 1979, \apj, 228, 939

\bibitem[{{Caswell} {et~al.}(1980){Caswell}, {Haynes}, {Milne}, \&
  {Wellington}}]{1980MNRAS.190..881C}
{Caswell}, J.~L., {Haynes}, R.~F., {Milne}, D.~K., \& {Wellington}, K.~J. 1980,
  \mnras, 190, 881

\bibitem[{{Chiotellis} {et~al.}(2012){Chiotellis}, {Schure}, \&
  {Vink}}]{2012A&A...537A.139C}
{Chiotellis}, A., {Schure}, K.~M., \& {Vink}, J. 2012, \aap, 537, A139

\bibitem[{{Clark} {et~al.}(1975){Clark}, {Caswell}, \&
  {Green}}]{1975AuJPA..37....1C}
{Clark}, D.~H., {Caswell}, J.~L., \& {Green}, A.~J. 1975, Australian Journal of
  Physics Astrophysical Supplement, 37, 1

\bibitem[{Cordes \& Lazio(2003)}]{cordes2003ne2001i}
Cordes, J.~M. \& Lazio, T. J.~W. 2003, NE2001.I. A New Model for the Galactic
  Distribution of Free Electrons and its Fluctuations

\bibitem[{{Dame} {et~al.}(2001){Dame}, {Hartmann}, \&
  {Thaddeus}}]{2001ApJ...547..792D}
{Dame}, T.~M., {Hartmann}, D., \& {Thaddeus}, P. 2001, \apj, 547, 792

\bibitem[{Ebeling {et~al.}(2006)Ebeling, White, \&
  Rangarajan}]{10.1111/j.1365-2966.2006.10135.x}
Ebeling, H., White, D.~A., \& Rangarajan, F. V.~N. 2006, Monthly Notices of the
  Royal Astronomical Society, 368, 65

\bibitem[{{Ferrand} \& {Safi-Harb}(2012)}]{2012AdSpR..49.1313F}
{Ferrand}, G. \& {Safi-Harb}, S. 2012, Advances in Space Research, 49, 1313

\bibitem[{{Finkbeiner}(2003)}]{2003ApJS..146..407F}
{Finkbeiner}, D.~P. 2003, \apjs, 146, 407

\bibitem[{{Foight} {et~al.}(2016){Foight}, {G{\"u}ver}, {{\"O}zel}, \&
  {Slane}}]{2016ApJ...826...66F}
{Foight}, D.~R., {G{\"u}ver}, T., {{\"O}zel}, F., \& {Slane}, P.~O. 2016, \apj,
  826, 66

\bibitem[{{Green} {et~al.}(1997){Green}, {Frail}, {Goss}, \&
  {Otrupcek}}]{1997AJ....114.2058G}
{Green}, A.~J., {Frail}, D.~A., {Goss}, W.~M., \& {Otrupcek}, R. 1997, \aj,
  114, 2058

\bibitem[{{Green}(2019)}]{2019JApA...40...36G}
{Green}, D.~A. 2019, Journal of Astrophysics and Astronomy, 40, 36

\bibitem[{{H.~E.~S.~S. Collaboration} {et~al.}(2018{\natexlab{a}}){H.~E.~S.~S.
  Collaboration}, {Abdalla}, {Abramowski}, {Aharonian}, {Ait Benkhali},
  {Ang{\"u}ner}, {Arakawa}, {Arrieta}, {Aubert}, {Backes}, {Balzer}, {Barnard},
  {Becherini}, {Becker Tjus}, {Berge}, {Bernhard}, {Bernl{\"o}hr}, {Blackwell},
  {B{\"o}ttcher}, {Boisson}, {Bolmont}, {Bonnefoy}, {Bordas}, {Bregeon},
  {Brun}, {Brun}, {Bryan}, {B{\"u}chele}, {Bulik}, {Capasso}, {Caroff},
  {Carosi}, {Casanova}, {Cerruti}, {Chakraborty}, {Chaves}, {Chen},
  {Chevalier}, {Colafrancesco}, {Condon}, {Conrad}, {Davids}, {Decock}, {Deil},
  {Devin}, {deWilt}, {Dirson}, {Djannati-Ata{\"\i}}, {Donath}, {Drury},
  {Dutson}, {Dyks}, {Edwards}, {Egberts}, {Emery}, {Ernenwein}, {Eschbach},
  {Farnier}, {Fegan}, {Fernandes}, {Fernandez}, {Fiasson}, {Fontaine}, {Funk},
  {F{\"u}{\ss}ling}, {Gabici}, {Gallant}, {Garrigoux}, {Gat{\'e}}, {Giavitto},
  {Giebels}, {Glawion}, {Glicenstein}, {Gottschall}, {Grondin}, {Hahn},
  {Haupt}, {Hawkes}, {Heinzelmann}, {Henri}, {Hermann}, {Hinton}, {Hofmann},
  {Hoischen}, {Holch}, {Holler}, {Horns}, {Ivascenko}, {Iwasaki},
  {Jacholkowska}, {Jamrozy}, {Jankowsky}, {Jankowsky}, {Jingo}, {Jouvin},
  {Jung-Richardt}, {Kastendieck}, {Katarzy{\'n}ski}, {Katsuragawa}, {Katz},
  {Kerszberg}, {Khangulyan}, {Kh{\'e}lifi}, {King}, {Klepser}, {Klochkov},
  {Klu{\'z}niak}, {Komin}, {Kosack}, {Krakau}, {Kraus}, {Kr{\"u}ger}, {Laffon},
  {Lamanna}, {Lau}, {Lees}, {Lefaucheur}, {Lemi{\`e}re}, {Lemoine-Goumard},
  {Lenain}, {Leser}, {Lohse}, {Lorentz}, {Liu}, {L{\'o}pez-Coto}, {Lypova},
  {Malyshev}, {Marandon}, {Marcowith}, {Mariaud}, {Marx}, {Maurin}, {Maxted},
  {Mayer}, {Meintjes}, {Meyer}, {Mitchell}, {Moderski}, {Mohamed}, {Mohrmann},
  {Mor{\r{a}}}, {Moulin}, {Murach}, {Nakashima}, {de Naurois}, {Ndiyavala},
  {Niederwanger}, {Niemiec}, {Oakes}, {O'Brien}, {Odaka}, {Ohm}, {Ostrowski},
  {Oya}, {Padovani}, {Panter}, {Parsons}, {Pekeur}, {Pelletier}, {Perennes},
  {Petrucci}, {Peyaud}, {Piel}, {Pita}, {Poireau}, {Poon}, {Prokhorov},
  {Prokoph}, {P{\"u}hlhofer}, {Punch}, {Quirrenbach}, {Raab}, {Rauth},
  {Reimer}, {Reimer}, {Renaud}, {de los Reyes}, {Rieger}, {Rinchiuso},
  {Romoli}, {Rowell}, {Rudak}, {Rulten}, {Safi-Harb}, {Sahakian}, {Saito},
  {Sanchez}, {Santangelo}, {Sasaki}, {Schlickeiser}, {Sch{\"u}ssler}, {Schulz},
  {Schwanke}, {Schwemmer}, {Seglar-Arroyo}, {Settimo}, {Seyffert}, {Shafi},
  {Shilon}, {Shiningayamwe}, {Simoni}, {Sol}, {Spanier}, {Spir-Jacob},
  {Stawarz}, {Steenkamp}, {Stegmann}, {Steppa}, {Sushch}, {Takahashi},
  {Tavernet}, {Tavernier}, {Taylor}, {Terrier}, {Tibaldo}, {Tiziani},
  {Tluczykont}, {Trichard}, {Tsirou}, {Tsuji}, {Tuffs}, {Uchiyama}, {van der
  Walt}, {van Eldik}, {van Rensburg}, {van Soelen}, {Vasileiadis}, {Veh},
  {Venter}, {Viana}, {Vincent}, {Vink}, {Voisin}, {V{\"o}lk}, {Vuillaume},
  {Wadiasingh}, {Wagner}, {Wagner}, {Wagner}, {White}, {Wierzcholska},
  {Willmann}, {W{\"o}rnlein}, {Wouters}, {Yang}, {Zaborov}, {Zacharias},
  {Zanin}, {Zdziarski}, {Zech}, {Zefi}, {Ziegler}, {Zorn}, \&
  {{\.Z}ywucka}}]{2018A&A...612A...3H}
{H.~E.~S.~S. Collaboration}, {Abdalla}, H., {Abramowski}, A., {et~al.}
  2018{\natexlab{a}}, \aap, 612, A3

\bibitem[{{H.~E.~S.~S. Collaboration} {et~al.}(2018{\natexlab{b}}){H.~E.~S.~S.
  Collaboration}, {Abdalla}, {Abramowski}, {Aharonian}, {Ait Benkhali},
  {Ang{\"u}ner}, {Arakawa}, {Arrieta}, {Aubert}, {Backes}, {Balzer}, {Barnard},
  {Becherini}, {Becker Tjus}, {Berge}, {Bernhard}, {Bernl{\"o}hr}, {Blackwell},
  {B{\"o}ttcher}, {Boisson}, {Bolmont}, {Bonnefoy}, {Bordas}, {Bregeon},
  {Brun}, {Brun}, {Bryan}, {B{\"u}chele}, {Bulik}, {Capasso}, {Carrigan},
  {Caroff}, {Carosi}, {Casanova}, {Cerruti}, {Chakraborty}, {Chaves}, {Chen},
  {Chevalier}, {Colafrancesco}, {Condon}, {Conrad}, {Davids}, {Decock}, {Deil},
  {Devin}, {deWilt}, {Dirson}, {Djannati-Ata{\"\i}}, {Domainko}, {Donath},
  {Drury}, {Dutson}, {Dyks}, {Edwards}, {Egberts}, {Eger}, {Emery},
  {Ernenwein}, {Eschbach}, {Farnier}, {Fegan}, {Fernandes}, {Fiasson},
  {Fontaine}, {F{\"o}rster}, {Funk}, {F{\"u}{\ss}ling}, {Gabici}, {Gallant},
  {Garrigoux}, {Gast}, {Gat{\'e}}, {Giavitto}, {Giebels}, {Glawion},
  {Glicenstein}, {Gottschall}, {Grondin}, {Hahn}, {Haupt}, {Hawkes},
  {Heinzelmann}, {Henri}, {Hermann}, {Hinton}, {Hofmann}, {Hoischen}, {Holch},
  {Holler}, {Horns}, {Ivascenko}, {Iwasaki}, {Jacholkowska}, {Jamrozy},
  {Jankowsky}, {Jankowsky}, {Jingo}, {Jouvin}, {Jung-Richardt}, {Kastendieck},
  {Katarzy{\'n}ski}, {Katsuragawa}, {Katz}, {Kerszberg}, {Khangulyan},
  {Kh{\'e}lifi}, {King}, {Klepser}, {Klochkov}, {Klu{\'z}niak}, {Komin},
  {Kosack}, {Krakau}, {Kraus}, {Kr{\"u}ger}, {Laffon}, {Lamanna}, {Lau},
  {Lees}, {Lefaucheur}, {Lemi{\`e}re}, {Lemoine-Goumard}, {Lenain}, {Leser},
  {Lohse}, {Lorentz}, {Liu}, {L{\'o}pez-Coto}, {Lypova}, {Marandon},
  {Malyshev}, {Marcowith}, {Mariaud}, {Marx}, {Maurin}, {Maxted}, {Mayer},
  {Meintjes}, {Meyer}, {Mitchell}, {Moderski}, {Mohamed}, {Mohrmann},
  {Mor{\r{a}}}, {Moulin}, {Murach}, {Nakashima}, {de Naurois}, {Ndiyavala},
  {Niederwanger}, {Niemiec}, {Oakes}, {O'Brien}, {Odaka}, {Ohm}, {Ostrowski},
  {Oya}, {Padovani}, {Panter}, {Parsons}, {Paz Arribas}, {Pekeur}, {Pelletier},
  {Perennes}, {Petrucci}, {Peyaud}, {Piel}, {Pita}, {Poireau}, {Poon},
  {Prokhorov}, {Prokoph}, {P{\"u}hlhofer}, {Punch}, {Quirrenbach}, {Raab},
  {Rauth}, {Reimer}, {Reimer}, {Renaud}, {de los Reyes}, {Rieger}, {Rinchiuso},
  {Romoli}, {Rowell}, {Rudak}, {Rulten}, {Safi-Harb}, {Sahakian}, {Saito},
  {Sanchez}, {Santangelo}, {Sasaki}, {Schandri}, {Schlickeiser},
  {Sch{\"u}ssler}, {Schulz}, {Schwanke}, {Schwemmer}, {Seglar-Arroyo},
  {Settimo}, {Seyffert}, {Shafi}, {Shilon}, {Shiningayamwe}, {Simoni}, {Sol},
  {Spanier}, {Spir-Jacob}, {Stawarz}, {Steenkamp}, {Stegmann}, {Steppa},
  {Sushch}, {Takahashi}, {Tavernet}, {Tavernier}, {Taylor}, {Terrier},
  {Tibaldo}, {Tiziani}, {Tluczykont}, {Trichard}, {Tsirou}, {Tsuji}, {Tuffs},
  {Uchiyama}, {van der Walt}, {van Eldik}, {van Rensburg}, {van Soelen},
  {Vasileiadis}, {Veh}, {Venter}, {Viana}, {Vincent}, {Vink}, {Voisin},
  {V{\"o}lk}, {Vuillaume}, {Wadiasingh}, {Wagner}, {Wagner}, {Wagner}, {White},
  {Wierzcholska}, {Willmann}, {W{\"o}rnlein}, {Wouters}, {Yang}, {Zaborov},
  {Zacharias}, {Zanin}, {Zdziarski}, {Zech}, {Zefi}, {Ziegler}, {Zorn}, \&
  {{\.Z}ywucka}}]{2018A&A...612A...1H}
{H.~E.~S.~S. Collaboration}, {Abdalla}, H., {Abramowski}, A., {et~al.}
  2018{\natexlab{b}}, \aap, 612, A1

\bibitem[{{H.E.S.S. Collaboration} {et~al.}(2018){H.E.S.S. Collaboration},
  {Abdalla}, {Abramowski}, {Aharonian}, {Ait Benkhali}, {Akhperjanian},
  {Andersson}, {Ang{\"u}ner}, {Arakawa}, {Arrieta}, {Aubert}, {Backes},
  {Balzer}, {Barnard}, {Becherini}, {Becker Tjus}, {Berge}, {Bernhard},
  {Bernl{\"o}hr}, {Blackwell}, {B{\"o}ttcher}, {Boisson}, {Bolmont},
  {Bonnefoy}, {Bordas}, {Bregeon}, {Brun}, {Brun}, {Bryan}, {B{\"u}chele},
  {Bulik}, {Capasso}, {Carr}, {Casanova}, {Cerruti}, {Chakraborty}, {Chaves},
  {Chen}, {Chevalier}, {Coffaro}, {Colafrancesco}, {Cologna}, {Condon},
  {Conrad}, {Cui}, {Davids}, {Decock}, {Degrange}, {Deil}, {Devin}, {deWilt},
  {Dirson}, {Djannati-Ata{\"\i}}, {Domainko}, {Donath}, {Drury}, {Dutson},
  {Dyks}, {Edwards}, {Egberts}, {Eger}, {Ernenwein}, {Eschbach}, {Farnier},
  {Fegan}, {Fernandes}, {Fiasson}, {Fontaine}, {F{\"o}rster}, {Funk},
  {F{\"u}{\ss}ling}, {Gabici}, {Gajdus}, {Gallant}, {Garrigoux}, {Giavitto},
  {Giebels}, {Glicenstein}, {Gottschall}, {Goyal}, {Grondin}, {Hahn}, {Haupt},
  {Hawkes}, {Heinzelmann}, {Henri}, {Hermann}, {Hervet}, {Hinton}, {Hofmann},
  {Hoischen}, {Holch}, {Holler}, {Horns}, {Ivascenko}, {Iwasaki},
  {Jacholkowska}, {Jamrozy}, {Janiak}, {Jankowsky}, {Jankowsky}, {Jingo},
  {Jogler}, {Jouvin}, {Jung-Richardt}, {Kastendieck}, {Katarzy{\'n}ski},
  {Katsuragawa}, {Katz}, {Kerszberg}, {Khangulyan}, {Kh{\'e}lifi}, {King},
  {Klepser}, {Klochkov}, {Klu{\'z}niak}, {Kolitzus}, {Komin}, {Kosack},
  {Krakau}, {Kraus}, {Kr{\"u}ger}, {Laffon}, {Lamanna}, {Lau}, {Lees},
  {Lefaucheur}, {Lefranc}, {Lemi{\`e}re}, {Lemoine-Goumard}, {Lenain}, {Leser},
  {Lohse}, {Lorentz}, {Liu}, {L{\'o}pez-Coto}, {Lypova}, {Marandon},
  {Marcowith}, {Mariaud}, {Marx}, {Maurin}, {Maxted}, {Mayer}, {Meintjes},
  {Meyer}, {Mitchell}, {Moderski}, {Mohamed}, {Mohrmann}, {Mor{\r{a}}},
  {Moulin}, {Murach}, {Nakashima}, {de Naurois}, {Niederwanger}, {Niemiec},
  {Oakes}, {O'Brien}, {Odaka}, {{\"O}ttl}, {Ohm}, {Ostrowski}, {Oya},
  {Padovani}, {Panter}, {Parsons}, {Pekeur}, {Pelletier}, {Perennes},
  {Petrucci}, {Peyaud}, {Piel}, {Pita}, {Poon}, {Prokhorov}, {Prokoph},
  {P{\"u}hlhofer}, {Punch}, {Quirrenbach}, {Raab}, {Reimer}, {Reimer},
  {Renaud}, {de los Reyes}, {Richter}, {Rieger}, {Romoli}, {Rowell}, {Rudak},
  {Rulten}, {Sahakian}, {Saito}, {Salek}, {Sanchez}, {Santangelo}, {Sasaki},
  {Schlickeiser}, {Sch{\"u}ssler}, {Schulz}, {Schwanke}, {Schwemmer},
  {Seglar-Arroyo}, {Settimo}, {Seyffert}, {Shafi}, {Shilon}, {Simoni}, {Sol},
  {Spanier}, {Spengler}, {Spies}, {Stawarz}, {Steenkamp}, {Stegmann}, {Stycz},
  {Sushch}, {Takahashi}, {Tavernet}, {Tavernier}, {Taylor}, {Terrier},
  {Tibaldo}, {Tiziani}, {Tluczykont}, {Trichard}, {Tsuji}, {Tuffs}, {Uchiyama},
  {van der Walt}, {van Eldik}, {van Rensburg}, {van Soelen}, {Vasileiadis},
  {Veh}, {Venter}, {Viana}, {Vincent}, {Vink}, {Voisin}, {V{\"o}lk},
  {Vuillaume}, {Wadiasingh}, {Wagner}, {Wagner}, {Wagner}, {White},
  {Wierzcholska}, {Willmann}, {W{\"o}rnlein}, {Wouters}, {Yang}, {Zabalza},
  {Zaborov}, {Zacharias}, {Zanin}, {Zdziarski}, {Zech}, {Zefi}, {Ziegler},
  {{\.Z}ywucka}, {Bamba}, {Fukui}, {Sano}, \& {Yoshiike}}]{2018A&A...612A...8H}
{H.E.S.S. Collaboration}, {Abdalla}, H., {Abramowski}, A., {et~al.} 2018, \aap,
  612, A8

\bibitem[{{Khabibullin} {et~al.}(2024){Khabibullin}, {Churazov}, {Chugai},
  {Bykov}, {Sunyaev}, {Utrobin}, {Zinchenko}, {Michailidis}, {Puehlhofer},
  {Becker}, {Freyberg}, {Merloni}, {Santangelo}, \&
  {Sasaki}}]{2024arXiv240117261K}
{Khabibullin}, I.~I., {Churazov}, E.~M., {Chugai}, N.~N., {et~al.} 2024,
  accepted to \aap, arXiv e-prints, arXiv:2401.17261

\bibitem[{{Koyama} {et~al.}(1995){Koyama}, {Petre}, {Gotthelf}, {Hwang},
  {Matsuura}, {Ozaki}, \& {Holt}}]{1995Natur.378..255K}
{Koyama}, K., {Petre}, R., {Gotthelf}, E.~V., {et~al.} 1995, \nat, 378, 255

\bibitem[{{Lallement} {et~al.}(2022){Lallement}, {Vergely}, {Babusiaux}, \&
  {Cox}}]{2022A&A...661A.147L}
{Lallement}, R., {Vergely}, J.~L., {Babusiaux}, C., \& {Cox}, N.~L.~J. 2022,
  \aap, 661, A147

\bibitem[{Leahy \& Williams(2017)}]{Leahy_2017}
Leahy, D.~A. \& Williams, J.~E. 2017, The Astronomical Journal, 153, 239

\bibitem[{{Mantovanini} {et~al.}(2024){Mantovanini}, {Becker}, {Khokhriakova},
  {Hurley-Walker}, {Anderson}, \& {Nicastro}}]{Mantovanini}
{Mantovanini}, S., {Becker}, W., {Khokhriakova}, A., {et~al.} 2024, submitted
  to \aap, arXiv e-prints, arXiv:2401.17294

\bibitem[{{Masai}(1984)}]{1984Ap&SS..98..367M}
{Masai}, K. 1984, \apss, 98, 367

\bibitem[{{Merloni} {et~al.}(2024){Merloni}, {Lamer}, {Liu}, {Ramos-Ceja},
  {Brunner}, {Bulbul}, {Dennerl}, {Doroshenko}, {Freyberg}, {Friedrich},
  {Gatuzz}, {Georgakakis}, {Haberl}, {Igo}, {Kreykenbohm}, {Liu}, {Maitra},
  {Malyali}, {Mayer}, {Nandra}, {Predehl}, {Robrade}, {Salvato}, {Sanders},
  {Stewart}, {Tub{\'\i}n-Arenas}, {Weber}, {Wilms}, {Arcodia}, {Artis},
  {Aschersleben}, {Avakyan}, {Aydar}, {Bahar}, {Balzer}, {Becker}, {Berger},
  {Boller}, {Bornemann}, {Br{\"u}ggen}, {Brusa}, {Buchner}, {Burwitz},
  {Camilloni}, {Clerc}, {Comparat}, {Coutinho}, {Czesla}, {Dannhauer},
  {Dauner}, {Dauser}, {Dietl}, {Dolag}, {Dwelly}, {Egg}, {Ehl}, {Freund},
  {Friedrich}, {Gaida}, {Garrel}, {Ghirardini}, {Gokus}, {Gr{\"u}nwald},
  {Grandis}, {Grotova}, {Gruen}, {Gueguen}, {H{\"a}mmerich}, {Hamaus},
  {Hasinger}, {Haubner}, {Homan}, {Ider Chitham}, {Joseph}, {Joyce},
  {K{\"o}nig}, {Kaltenbrunner}, {Khokhriakova}, {Kink}, {Kirsch}, {Kluge},
  {Knies}, {Krippendorf}, {Krumpe}, {Kurpas}, {Li}, {Liu}, {Locatelli},
  {Lorenz}, {M{\"u}ller}, {Magaudda}, {Mannes}, {McCall}, {Meidinger},
  {Michailidis}, {Migkas}, {Mu{\~n}oz-Giraldo}, {Musiimenta}, {Nguyen-Dang},
  {Ni}, {Olechowska}, {Ota}, {Pacaud}, {Pasini}, {Perinati}, {Pires},
  {Pommranz}, {Ponti}, {Poppenhaeger}, {P{\"u}hlhofer}, {Rau}, {Reh},
  {Reiprich}, {Roster}, {Saeedi}, {Santangelo}, {Sasaki}, {Schmitt},
  {Schneider}, {Schrabback}, {Schuster}, {Schwope}, {Seppi}, {Serim},
  {Shreeram}, {Sokolova-Lapa}, {Starck}, {Stelzer}, {Stierhof}, {Suleimanov},
  {Tenzer}, {Traulsen}, {Tr{\"u}mper}, {Tsuge}, {Urrutia}, {Veronica},
  {Waddell}, {Willer}, {Wolf}, {Yeung}, {Zainab}, {Zangrandi}, {Zhang},
  {Zhang}, \& {Zheng}}]{Merloni2023}
{Merloni}, A., {Lamer}, G., {Liu}, T., {et~al.} 2024, \aap, 682, A34

\bibitem[{{Merloni} {et~al.}(2012){Merloni}, {Predehl}, {Becker},
  {B{\"o}hringer}, {Boller}, {Brunner}, {Brusa}, {Dennerl}, {Freyberg},
  {Friedrich}, {Georgakakis}, {Haberl}, {Hasinger}, {Meidinger}, {Mohr},
  {Nandra}, {Rau}, {Reiprich}, {Robrade}, {Salvato}, {Santangelo}, {Sasaki},
  {Schwope}, {Wilms}, \& {German eROSITA Consortium}}]{2012arXiv1209.3114M}
{Merloni}, A., {Predehl}, P., {Becker}, W., {et~al.} 2012, eROSITA science
  book, arXiv e-prints, arXiv:1209.3114

\bibitem[{{Michailidis} {et~al.}(2024{\natexlab{a}}){Michailidis},
  {P{\"u}hlhofer}, {Becker}, {Freyberg}, {Merloni}, {Santangelo}, {Sasaki},
  {Bykov}, {Chugai}, {Churazov}, {Khabibullin}, {Sunyaev}, {Utrobin}, \&
  {Zinchenko}}]{2024arXiv240117312M}
{Michailidis}, M., {P{\"u}hlhofer}, G., {Becker}, W., {et~al.}
  2024{\natexlab{a}}, accepted to \aap, arXiv e-prints, arXiv:2401.17312

\bibitem[{{Michailidis} {et~al.}(2024{\natexlab{b}}){Michailidis},
  {P{\"u}hlhofer}, {Santangelo}, {Becker}, \& {Sasaki}}]{2024A&A...685A..23M}
{Michailidis}, M., {P{\"u}hlhofer}, G., {Santangelo}, A., {Becker}, W., \&
  {Sasaki}, M. 2024{\natexlab{b}}, \aap, 685, A23

\bibitem[{{Nolan} {et~al.}(2012){Nolan}, {Abdo}, {Ackermann}, {Ajello},
  {Allafort}, {Antolini}, {Atwood}, {Axelsson}, {Baldini}, {Ballet},
  {Barbiellini}, {Bastieri}, {Bechtol}, {Belfiore}, {Bellazzini}, {Berenji},
  {Bignami}, {Blandford}, {Bloom}, {Bonamente}, {Bonnell}, {Borgland},
  {Bottacini}, {Bouvier}, {Brandt}, {Bregeon}, {Brigida}, {Bruel}, {Buehler},
  {Burnett}, {Buson}, {Caliandro}, {Cameron}, {Campana}, {Ca{\~n}adas},
  {Cannon}, {Caraveo}, {Casandjian}, {Cavazzuti}, {Ceccanti}, {Cecchi},
  {{\c{C}}elik}, {Charles}, {Chekhtman}, {Cheung}, {Chiang}, {Chipaux},
  {Ciprini}, {Claus}, {Cohen-Tanugi}, {Cominsky}, {Conrad}, {Corbet}, {Cutini},
  {D'Ammando}, {Davis}, {de Angelis}, {DeCesar}, {DeKlotz}, {De Luca}, {den
  Hartog}, {de Palma}, {Dermer}, {Digel}, {Silva}, {Drell}, {Drlica-Wagner},
  {Dubois}, {Dumora}, {Enoto}, {Escande}, {Fabiani}, {Falletti}, {Favuzzi},
  {Fegan}, {Ferrara}, {Focke}, {Fortin}, {Frailis}, {Fukazawa}, {Funk},
  {Fusco}, {Gargano}, {Gasparrini}, {Gehrels}, {Germani}, {Giebels},
  {Giglietto}, {Giommi}, {Giordano}, {Giroletti}, {Glanzman}, {Godfrey},
  {Grenier}, {Grondin}, {Grove}, {Guillemot}, {Guiriec}, {Gustafsson},
  {Hadasch}, {Hanabata}, {Harding}, {Hayashida}, {Hays}, {Hill}, {Horan},
  {Hou}, {Hughes}, {Iafrate}, {Itoh}, {J{\'o}hannesson}, {Johnson}, {Johnson},
  {Johnson}, {Johnson}, {Kamae}, {Katagiri}, {Kataoka}, {Katsuta}, {Kawai},
  {Kerr}, {Kn{\"o}dlseder}, {Kocevski}, {Kuss}, {Lande}, {Landriu},
  {Latronico}, {Lemoine-Goumard}, {Lionetto}, {Llena Garde}, {Longo},
  {Loparco}, {Lott}, {Lovellette}, {Lubrano}, {Madejski}, {Marelli}, {Massaro},
  {Mazziotta}, {McConville}, {McEnery}, {Mehault}, {Michelson}, {Minuti},
  {Mitthumsiri}, {Mizuno}, {Moiseev}, {Mongelli}, {Monte}, {Monzani},
  {Morselli}, {Moskalenko}, {Murgia}, {Nakamori}, {Naumann-Godo}, {Norris},
  {Nuss}, {Nymark}, {Ohno}, {Ohsugi}, {Okumura}, {Omodei}, {Orlando}, {Ormes},
  {Ozaki}, {Paneque}, {Panetta}, {Parent}, {Perkins}, {Pesce-Rollins},
  {Pierbattista}, {Pinchera}, {Piron}, {Pivato}, {Porter}, {Racusin},
  {Rain{\`o}}, {Rando}, {Razzano}, {Razzaque}, {Reimer}, {Reimer}, {Reposeur},
  {Ritz}, {Rochester}, {Romani}, {Roth}, {Rousseau}, {Ryde}, {Sadrozinski},
  {Salvetti}, {Sanchez}, {Saz Parkinson}, {Sbarra}, {Scargle}, {Schalk},
  {Sgr{\`o}}, {Shaw}, {Shrader}, {Siskind}, {Smith}, {Spandre}, {Spinelli},
  {Stephens}, {Strickman}, {Suson}, {Tajima}, {Takahashi}, {Takahashi},
  {Tanaka}, {Thayer}, {Thayer}, {Thompson}, {Tibaldo}, {Tibolla}, {Tinebra},
  {Tinivella}, {Torres}, {Tosti}, {Troja}, {Uchiyama}, {Vandenbroucke}, {Van
  Etten}, {Van Klaveren}, {Vasileiou}, {Vianello}, {Vitale}, {Waite},
  {Wallace}, {Wang}, {Werner}, {Winer}, {Wood}, {Wood}, {Wood}, {Yang}, \&
  {Zimmer}}]{2012ApJS..199...31N}
{Nolan}, P.~L., {Abdo}, A.~A., {Ackermann}, M., {et~al.} 2012, \apjs, 199, 31

\bibitem[{{Pavlovi{\'c}} {et~al.}(2013){Pavlovi{\'c}}, {Uro{\v{s}}evi{\'c}},
  {Vukoti{\'c}}, {Arbutina}, \& {G{\"o}ker}}]{2013ApJS..204....4P}
{Pavlovi{\'c}}, M.~Z., {Uro{\v{s}}evi{\'c}}, D., {Vukoti{\'c}}, B., {Arbutina},
  B., \& {G{\"o}ker}, {\"U}.~D. 2013, \apjs, 204, 4

\bibitem[{Pearson(1900)}]{Pearson1900}
Pearson, K. 1900, The London, Edinburgh, and Dublin Philosophical Magazine and
  Journal of Science, 50, 157

\bibitem[{{Predehl} {et~al.}(2021){Predehl}, {Andritschke}, {Arefiev},
  {Babyshkin}, {Batanov}, {Becker}, {B{\"o}hringer}, {Bogomolov}, {Boller},
  {Borm}, {Bornemann}, {Br{\"a}uninger}, {Br{\"u}ggen}, {Brunner}, {Brusa},
  {Bulbul}, {Buntov}, {Burwitz}, {Burkert}, {Clerc}, {Churazov}, {Coutinho},
  {Dauser}, {Dennerl}, {Doroshenko}, {Eder}, {Emberger}, {Eraerds},
  {Finoguenov}, {Freyberg}, {Friedrich}, {Friedrich}, {F{\"u}rmetz},
  {Georgakakis}, {Gilfanov}, {Granato}, {Grossberger}, {Gueguen}, {Gureev},
  {Haberl}, {H{\"a}lker}, {Hartner}, {Hasinger}, {Huber}, {Ji}, {Kienlin},
  {Kink}, {Korotkov}, {Kreykenbohm}, {Lamer}, {Lomakin}, {Lapshov}, {Liu},
  {Maitra}, {Meidinger}, {Menz}, {Merloni}, {Mernik}, {Mican}, {Mohr},
  {M{\"u}ller}, {Nandra}, {Nazarov}, {Pacaud}, {Pavlinsky}, {Perinati},
  {Pfeffermann}, {Pietschner}, {Ramos-Ceja}, {Rau}, {Reiffers}, {Reiprich},
  {Robrade}, {Salvato}, {Sanders}, {Santangelo}, {Sasaki}, {Scheuerle},
  {Schmid}, {Schmitt}, {Schwope}, {Shirshakov}, {Steinmetz}, {Stewart},
  {Str{\"u}der}, {Sunyaev}, {Tenzer}, {Tiedemann}, {Tr{\"u}mper}, {Voron},
  {Weber}, {Wilms}, \& {Yaroshenko}}]{2021A&A...647A...1P}
{Predehl}, P., {Andritschke}, R., {Arefiev}, V., {et~al.} 2021, \aap, 647, A1

\bibitem[{{Sunyaev} {et~al.}(2021){Sunyaev}, {Arefiev}, {Babyshkin},
  {Bogomolov}, {Borisov}, {Buntov}, {Brunner}, {Burenin}, {Churazov},
  {Coutinho}, {Eder}, {Eismont}, {Freyberg}, {Gilfanov}, {Gureyev}, {Hasinger},
  {Khabibullin}, {Kolmykov}, {Komovkin}, {Krivonos}, {Lapshov}, {Levin},
  {Lomakin}, {Lutovinov}, {Medvedev}, {Merloni}, {Mernik}, {Mikhailov},
  {Molodtsov}, {Mzhelsky}, {M{\"u}ller}, {Nandra}, {Nazarov}, {Pavlinsky},
  {Poghodin}, {Predehl}, {Robrade}, {Sazonov}, {Scheuerle}, {Shirshakov},
  {Tkachenko}, \& {Voron}}]{2021A&A...656A.132S}
{Sunyaev}, R., {Arefiev}, V., {Babyshkin}, V., {et~al.} 2021, \aap, 656, A132

\bibitem[{{Wang} {et~al.}(2020){Wang}, {Zhang}, {Jiang}, {Zhao}, {Chen},
  {Chen}, {Gao}, \& {Liu}}]{2020A&A...639A..72W}
{Wang}, S., {Zhang}, C., {Jiang}, B., {et~al.} 2020, \aap, 639, A72

\bibitem[{{Whiteoak} \& {Green}(1996)}]{1996A&AS..118..329W}
{Whiteoak}, J.~B.~Z. \& {Green}, A.~J. 1996, \aaps, 118, 329

\bibitem[{{Wilms} {et~al.}(2000){Wilms}, {Allen}, \&
  {McCray}}]{2000ApJ...542..914W}
{Wilms}, J., {Allen}, A., \& {McCray}, R. 2000, \apj, 542, 914

\bibitem[{Yamaguchi {et~al.}(2011)Yamaguchi, Koyama, \&
  Uchida}]{10.1093/pasj/63.sp3.S837}
Yamaguchi, H., Koyama, K., \& Uchida, H. 2011, Publications of the Astronomical
  Society of Japan, 63, S837

\bibitem[{{Yao} {et~al.}(2017){Yao}, {Manchester}, \&
  {Wang}}]{2017ApJ...835...29Y}
{Yao}, J.~M., {Manchester}, R.~N., \& {Wang}, N. 2017, \apj, 835, 29

\bibitem[{{Yasuda} \& {Lee}(2019)}]{2019ApJ...876...27Y}
{Yasuda}, H. \& {Lee}, S.-H. 2019, \apj, 876, 27

\end{thebibliography}

\bibliographystyle{aa} 

\end{document}